\documentclass[11p]{article}
\pdfoutput=1
\usepackage[margin=1in]{geometry}
\usepackage[nodisplayskipstretch]{setspace}
\usepackage{microtype}
\usepackage{fullpage,enumitem,amssymb,amsmath,xcolor,cancel,graphicx,url}
\usepackage{indentfirst}
\usepackage{amsthm}
\usepackage{booktabs}
\usepackage{siunitx}
\usepackage[table]{colortbl}
\usepackage{float}
\usepackage{blkarray}
\usepackage[colorlinks=true, allcolors=blue]{hyperref}
\usepackage[numbers]{natbib}
\usepackage{mathtools}  
\usepackage{subcaption}
\usepackage{bbm}
\usepackage{tikz}
\usetikzlibrary{positioning,calc,positioning}
\usepackage{xcolor}
\usepackage{authblk}

\usepackage{algorithm}
\usepackage{algorithmic}

\makeatletter
\newcommand{\ALOOP}[1]{\ALC@it\algorithmicloop\ #1%
  \begin{ALC@loop}}
\newcommand{\ENDALOOP}{\end{ALC@loop}\ALC@it\algorithmicendloop}

\makeatother

\begin{document}

\title{\bf Bayesian estimation of clustered dependence structures in functional neuroconnectivity}
\author[1]{Hyoshin Kim}
\author[1]{Sujit K. Ghosh}
\author[2]{Adriana Di Martino}
\author[1]{Emily C. Hector \thanks{Hector was supported by a grant from the National Science Foundation (DMS2152887) and a Faculty Research and Professional Development Award from North Carolina State University.}}
\affil[1]{Department of Statistics, North Carolina State University}
\affil[2]{Autism Center, Child Mind Institute}
\date{}

\maketitle

\begin{abstract}
\noindent Motivated by the need to model the dependence between regions of interest in functional neuroconnectivity for efficient inference, we propose a new sampling-based Bayesian clustering approach for covariance structures of high-dimensional Gaussian outcomes. The key technique is based on a Dirichlet process that clusters covariance sub-matrices into independent groups of outcomes, thereby naturally inducing sparsity in the whole brain connectivity matrix. A new split-merge algorithm is employed to achieve convergence of the Markov chain that is shown empirically to recover both uniform and Dirichlet partitions with high accuracy. We investigate the empirical performance of the proposed method through extensive simulations. Finally, the proposed approach is used to group regions of interest into functionally independent groups in the Autism Brain Imaging Data Exchange participants with autism spectrum disorder and and co-occurring attention-deficit/hyperactivity disorder. 
\end{abstract}

\noindent%

\section{Introduction}\label{sec:intro}

Analyzing functional connectivity in the brain is a crucial task in the study of Autism Spectrum Disorder (ASD). The specific patterns of hyper- and hypoconnectivity between brain regions of interest (ROIs) of autistic versus neurotypical individuals remains poorly understood, with inconsistent findings regarding abnormal functional connectivity in ASD reported in existing neuroimaging studies \citep{Hulletal2016}. We consider the average resting-state fMRI (rfMRI) time series across voxels in each of $M$ ROIs for individuals $i\in \{1, \ldots, n\}$.  
The individual-specific functional connectivity between $M$ ROIs is estimated by computing the sample correlation between the $M$ averaged rfMRI time series, treating the time series points as independent replicates. In whole-brain analyses, this connectivity matrix is used to investigate patterns of coactivation of ROIs across disease status.

The computational complexity and Type-I error multiplicity associated with the connectivity matrix, which contains $O(M^2)$ entries for each individual, often leads studies to use coarse parcellations with larger ROIs, such as the Harvard-Oxford atlas with $M=111$ or the Craddock 200 atlas with $M=200$ \citep{Haetal2015, Ghiassianetal2016}. Studies aiming to investigate brain functional connectivity at a finer scale typically focus on specific sets of brain regions \citep{Beeretal2019, Lynchetal2013}, thereby limiting the scope of analyses and scientific discoveries. Moreover, multiple analyses focusing on different sets of brain regions run the risk of type-I error inflation, but multiple comparison adjustments are overly conservative when ROIs are correlated. While whole-brain voxel- or surface-wise analyses are increasingly used, they are computationally expensive and require large scale datasets. There is thus a need for statistical methodology that learns general sparsity patterns in the connectivity matrix of fine brain parcellations and provides appropriate quantification of uncertainty.

Motivated by this need, we aim to identify independent groups of correlated ROIs. Due to the large number of voxels in each ROI, the $M$ averaged rfMRI time series are approximately Gaussian owing to the Central Limit Theorem approximation. Our underlying assumption is that the $M$ outcomes naturally belong to independent groups, leading to a clustering of the covariance sub-matrices. In particular, we assume that the true covariance matrix is a permutation of a block-diagonal matrix composed of structured covariance sub-matrices. This construction ensures that the covariance matrix is positive-definite, capable of accommodating any arbitrary ordering of the ROIs, and flexible in capturing general sparsity patterns. We propose a sampling based Bayesian clustering approach that partitions the covariance matrix of a high-dimensional Gaussian outcome into structured sub-matrices. The independent groups of ROIs identified by our approach can be examined separately in downstream analyses without the fear of loss of statistical power. In Section \ref{sec:data}, our method identifies groups of correlated ROIs in individuals diagnosed with Autism Spectrum Disorder (ASD) and attention-deficit/hyperactivity disorder (ADHD) using rfMRI data from the Autism Brain Imaging Data Exchange (ABIDE) \citep{DiMartinoetal2014} preprocessed repository \citep{craddock2013neuro}. In this dataset, we use $M=1000$ ROIs based on the hierarchical multi-resolution parcellation of \cite{Schaeferetal2018}, defining a high-dimensional setting where the number of independent observations, denoted by $N$, can be comparable to the number of ROIs. 

The modeling of high-dimensional covariance matrices has received considerable attention in the literature. Existing methods that assume a group structure for the outcomes typically assume the group structure is predetermined or \emph{a priori} known \citep{Lietal2015, Luoetal2020}. Examples in spatial settings include  spectral methods \citep{Fuentes2007}, tapered covariance functions \citep{Kaufmanetal2008, Stein2013} and low-rank approximations \citep{Banerjeeetal2008, Nychkaetal2015}.  In contrast, our objective is to learn the group structure through clustering with distinct sets of covariance parameters for each group. Another line of research focused on modeling high-dimensional covariance matrices assumes a general sparsity pattern for the covariance or its inverse, the precision matrix. Bayesian methods in this field use prior distributions that assign zero probability mass to zero entries while concentrating mass on non-zero entries. Examples include the G-Wishart prior \citep{Roverato2002, MohammadiWit2015}, the Laplace prior \citep{Khondkeretal2013, Wang2012}, and the spike-and-slab prior \citep{Samantaetal2022, Deshpandeetal2019}. While these methods effectively handle general sparsity patterns, they are more suitable for identifying pairwise relationships between outcomes rather than discovering independent groups of correlated outcomes.

We propose a new Dirichlet process prior for clustering of covariance parameters \citep{Ferguson1973}. To enhance the mixing of cluster components within a Markov chain based sampling procedure, we present a new Metropolis-Hastings split-merge algorithm called the \emph{Head-Tail Split Merge (HTSM)} algorithm. In various applications, split-merge samplers are used alongside Gibbs samplers to improve mixing in models using the Dirichlet process, particularly the Dirichlet process mixture model. The state-of-the-art methods in this domain are the Metropolis-Hastings split-merge samplers proposed by \cite{Dahl2003} and \cite{JainNeal2004}. Other related approaches include the Metropolis-Hastings Smart-Dumb/Dumb-Smart sampler of \cite{WangRussell2015}, which combines smart split moves with dumb merge moves or vice versa. The Particle Gibbs Split-Merge sampler introduced by \cite{Alexandreetal2017} avoids the need for complex acceptance ratios in Metropolis-Hastings algorithms. Additionally, \cite{ChangFisher2013, williamsonetal2013} focus on parallelization and distribution of Markov Chain Monte Carlo (MCMC) methods applicable to split-merge samplers.

Although these existing split-merge samplers are effective, they are primarily designed for clustering in Dirichlet process mixture models. Typically, such models involve a small number of clusters $J$ and allow for a large dimension $M$, resulting in larger-sized clusters due to the nature of mixture models. In contrast, our focus is on clustering covariance sub-matrices, where both $M$ and $J$ can be large, allowing for a larger number of smaller-sized clusters, including clusters consisting of a single ROI. To address this scenario, we propose the HTSM algorithm, which can handle these settings effectively. The HTSM algorithm incorporates two types of proposals: a head split-merge proposal used in conjunction with the slice sampler introduced by \cite{Walker2014}, which is aimed at identifying large clusters, and a tail split-merge proposal employed to detect small clusters. In Section \ref{sec:sim}, we demonstrate empirically that the proposed algorithm can successfully recover both uniform and true Dirichlet partitions when both $M$ and $J$ are very large. We also show how to incorporate temporal or spatial information to construct the covariance sub-matrices.

The paper is organized as follows. In Section \ref{sec:model}, we formulate the Bayesian clustering model for covariance sub-matrices using the stick-breaking representation of the Dirichlet process prior. Section \ref{sec:samplerZ} focuses on the sampling of the cluster indices and explores the mixing properties associated with this sampling process. Additionally, this section introduces our proposed HTSM split-merge sampler designed to achieve convergence of the proposed MCMC clustering approach. The results of simulations are presented in Section \ref{sec:sim}, and an analysis of the ABIDE data is presented in Section \ref{sec:data}. Section \ref{sec:conc} concludes.

\section{A model for clustering covariance sub-matrices} \label{sec:model}

\subsection{A model for the independent groups of outcomes}

Let $y_{i}(\boldsymbol{s}_{m})$ be the $i$-th observed averaged rfMRI outcome at the centroid $\boldsymbol{s}_{m} \in \mathbb{R}^3$ of the $m$-th ROI, $m=1,\dots,M$, for $i=1,\dots,N$ independently. In our ABIDE data analysis (Section \ref{sec:data}), $N$ corresponds to the total number of (thinned) rfMRI time series for the set of individuals with ASD and ADHD. We define $\mathcal{S}$ as the set of $M$ locations, $\mathcal{S} = \{\boldsymbol{s}_{m}\}_{m=1}^{M}$, and define $\boldsymbol{y}_i(\mathcal{S}) = \{y_{i}(\boldsymbol{s}_{1}), \dots, y_{i}(\boldsymbol{s}_{M})\}$, where the order of the ROIs is arbitrary. Let $\{\boldsymbol{X}_{i}\}_{i=1}^{N}$ be $N$ $p$-dimensional vectors of covariates. We assume that $\mathbb{E}\{ \boldsymbol{y}_{i}(\mathcal{S}) \} = \boldsymbol{B}\boldsymbol{X}_{i}$ with $\boldsymbol{B} = (\boldsymbol{\beta}_1, \dots, \boldsymbol{\beta}_p) \in \mathbb{R}^{M \times p}$ a matrix of regression parameters, $\boldsymbol{\beta}_q \in \mathbb{R}^M$, $q=1, \ldots, p$. The variance $\mathbb{V}\{ \boldsymbol{y}_{i}(\mathcal{S}) \} = \boldsymbol{\boldsymbol{\Sigma}}(\mathcal{S}, \mathcal{S}; \boldsymbol{\theta})$ depends on a vector of covariance parameters $\boldsymbol{\theta}$ specified below, and does not depend on covariates $\boldsymbol{X}_i$. We assume that the outcomes independently follow $M$-variate Gaussian distributions given by
\begin{align*}
\boldsymbol{y}_i(\mathcal{S}) \sim \mathcal{N}_M \{ \boldsymbol{B} \boldsymbol{X}_{i}, \boldsymbol{\Sigma}(\mathcal{S}, \mathcal{S}; \boldsymbol{\theta}) \}.
\end{align*}
We assume that the outcomes in vector $\boldsymbol{y}_i(\mathcal{S})$ can be permuted to construct $\boldsymbol{y}_{i,\text{perm}}(\mathcal{S})=\{ \boldsymbol{y}_{i,\text{perm}}(\mathcal{S}_1), \ldots, \allowbreak \boldsymbol{y}_{i,\text{perm}}(\mathcal{S}_J)\}$ based on a partition of $\mathcal{S}$ into $J$ sets $\{\mathcal{S}_{1},\dots,\mathcal{S}_{J}\}$, where outcomes $\boldsymbol{y}_{i,\text{perm}}(\mathcal{S}_j)$ within a group $j$ are correlated with each other and outcomes $\boldsymbol{y}_{i,\text{perm}}(\mathcal{S}_j), \boldsymbol{y}_{i,\text{perm}}(\mathcal{S}_{j'})$ across groups $j,j'$ are independent. The independent groups of outcomes induce a block-diagonal structure on $\mathbb{V}\{\boldsymbol{y}_{i,\text{perm}}(\mathcal{S})\} = \boldsymbol{\Sigma}_{\text{perm}}(\mathcal{S}, \mathcal{S}; \boldsymbol{\theta})$ defined by 
\begin{align*}
    \boldsymbol{\Sigma}_{\text{perm}}(\mathcal{S}, \mathcal{S}; \boldsymbol{\theta}) = \text{Block-diag}\{\boldsymbol{\Sigma}_{\text{perm}}(\mathcal{S}_{1}, \mathcal{S}_{1}; \boldsymbol{\theta}_{1}), \dots, \boldsymbol{\Sigma}_{\text{perm}}(\mathcal{S}_{J}, \mathcal{S}_{J}; \boldsymbol{\theta}_{J})\}.
\end{align*}
The dimension of each diagonal block is denoted by $d_{j} = |\mathcal{S}_{j}|$, corresponding to the number of ROIs in the $j$-th partition set. Each diagonal block covariance matrix $\boldsymbol{\Sigma}_{\text{perm}}(\mathcal{S}_{j}, \mathcal{S}_{j}; \boldsymbol{\theta}_{j})$ is parameterized by a set of variance and correlation parameters $\boldsymbol{\theta}_{j} = (\sigma_{j}^2, \rho_{j})$, $j=1,\dots,J$, and we denote $\boldsymbol{\theta}=(\boldsymbol{\theta}_j)_{j=1}^J \in \mathbb{R}^{2J}$. We model each diagonal block covariance with
\begin{align*}
    \boldsymbol{\Sigma}_{\text{perm}}(\mathcal{S}_{j}, \mathcal{S}_{j}; \boldsymbol{\theta}_{j}) = \sigma_{j}^2 \boldsymbol{\Gamma}(\mathcal{S}_{j}, \mathcal{S}_{j}; \rho_{j}),
\end{align*}
where $\boldsymbol{\Gamma}(\mathcal{S}_{j}, \mathcal{S}_{j}; \rho_{j})$ is a pre-specified, stationary positive-definite correlation function. This modeling approach allows us to incorporate well-known positive-definite correlation functions, such as compound symmetry, generalized AR(1) \citep{MurrayHelms1990}, or Mat\'ern \citep{Matern1986} into our modeling framework.

\begin{figure}[ht!]
    \centering
    \begin{tikzpicture}
        \node (image) at (0,0) {
        \includegraphics[width=0.7\textwidth]{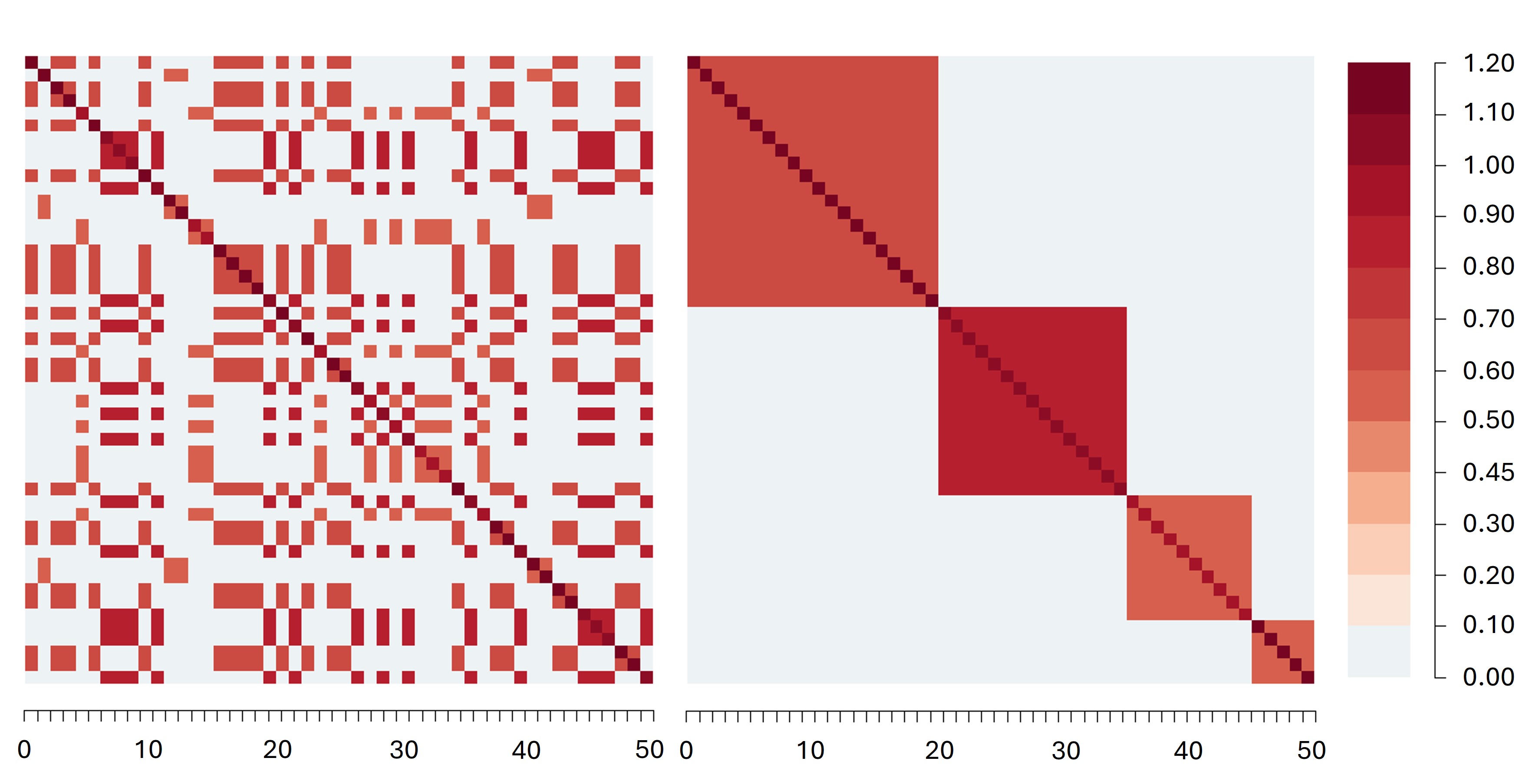}
        };
        \node[above,black] at (-3.2,2.6){\small $\boldsymbol{\Sigma}(\mathcal{S}, \mathcal{S}; \boldsymbol{\theta})$};
        \node[above,black] at (1.9,2.6){\small $\boldsymbol{\Sigma}_{\text{perm}}(\mathcal{S}, \mathcal{S}; \boldsymbol{\theta})$};
    \end{tikzpicture}
    \caption{Example $\boldsymbol{\Sigma}(\mathcal{S}, \mathcal{S}; \boldsymbol{\theta})$ (left) and $\boldsymbol{\Sigma}_{\text{perm}}(\mathcal{S}, \mathcal{S}; \boldsymbol{\theta})$ (right) for $M=50$.} \label{fig: example}
\end{figure}

We introduce the cluster index variable $Z_m \in \{1, \ldots, J\}$ to denote the cluster to which the outcome $y_i(\boldsymbol{s}_m)$ belongs. The cluster indices $\boldsymbol{Z} = (Z_1, \dots, Z_M)$ induce a partition $\{\mathcal{S}_1, \dots, \mathcal{S}_J\}$, where the number of clusters is determined by $J = \max_m Z_m$. We define a permutation $\pi: \{1, \dots, M\} \rightarrow \{1, \dots, M\}$ which rearranges the elements of $\boldsymbol{Z}$ in ascending order. The permutation matrix $\boldsymbol{P}_{\pi}$ is constructed by permuting the columns of the identity matrix: the $m$-th row of $\boldsymbol{P}_{\pi}$ is a standard basis row vector $\boldsymbol{e}_{\pi(m)} \in \mathbb{R}^{M}$ with zeros in all entries except for a single 1 located in the $\pi(m)$-th column. We thus define matrices $\boldsymbol{\pi}$ and $\boldsymbol{P}_{\pi}$ as 
\begin{align*}
    \boldsymbol{\pi} = 
    \begin{pmatrix}
    1&2&\dots&M\\
    \pi(1)&\pi(2)&\dots&\pi(M)
    \end{pmatrix}; \quad \boldsymbol{P}_{\pi} = 
    \begin{pmatrix}
    \boldsymbol{e}_{\pi(1)}\\
    \boldsymbol{e}_{\pi(2)}\\
    \vdots\\
    \boldsymbol{e}_{\pi(M)}
    \end{pmatrix}.
\end{align*}
The permutation matrix systematically recovers $\boldsymbol{\Sigma}(\mathcal{S}, \mathcal{S}; \boldsymbol{\theta})$ from $\boldsymbol{\Sigma}_{\text{perm}}(\mathcal{S}, \mathcal{S}; \boldsymbol{\theta})$ through
\begin{align*}
    \boldsymbol{\Sigma}(\mathcal{S}, \mathcal{S}; \boldsymbol{\theta}) = \boldsymbol{P}_{\pi}^{\top} \boldsymbol{\Sigma}_{\text{perm}}(\mathcal{S}, \mathcal{S}; \boldsymbol{\theta}) \boldsymbol{P}_{\pi},
\end{align*}
up to a permutation between blocks, allowing for arbitrary ordering of the diagonal blocks of $\boldsymbol{\Sigma}_{\text{perm}}$. In this fashion, $\boldsymbol{\Sigma}(\mathcal{S}, \mathcal{S}; \boldsymbol{\theta})$ is equivalent to $\boldsymbol{\Sigma}_{\text{perm}}(\mathcal{S}, \mathcal{S}; \boldsymbol{\theta})$ with its rows and columns permuted. We illustrate with a toy example of a covariance matrix with $M=50$ in Figure \ref{fig: example}. 

Our formulation offers several advantages. First, positive-definiteness of estimators of  $\boldsymbol{\Sigma}(\mathcal{S}, \mathcal{S}; \boldsymbol{\theta})$ is guaranteed without requiring additional steps or compromising computational efficiency. Second, our approach permits arbitrary ordering of the locations $\boldsymbol{s}_m$ so that no initial estimate of the partition is required. Third, sparsity estimated in the block-diagonal structure of $\boldsymbol{\Sigma}_{\text{perm}}(\mathcal{S}, \mathcal{S}; \boldsymbol{\theta})$ naturally induces sparsity in $\boldsymbol{\Sigma}(\mathcal{S}, \mathcal{S}; \boldsymbol{\theta})$, enabling our approach to flexibly capture different sparsity patterns. Finally, we allow $J$ to be determined by the data and not pre-fixed.

\subsection{Priors for the cluster indices and regression and covariance parameters}

We place a Dirichlet process prior on the distribution of the cluster parameters. Let $P \sim \text{DP}(\alpha, P_{0})$ denote a Dirichlet process prior distribution \citep{Ferguson1973} with a scale parameter $\alpha > 0$ and a prior probability $P_{0}$. We adopt the stick-breaking representation of the Dirichlet process prior \citep{Sethuraman1994} as follows: 
\begin{equation}
\begin{split}
     &P = \sum_{j=1}^{\infty} w_{j} \delta_{\boldsymbol{\theta}_{j}}, \quad w_{j} = V_{j} \prod_{l=1}^{j-1} (1-V_{l}),\\
     &\boldsymbol{\theta}_{j} \sim P_{0} \text{ for } j \in \mathbb{Z}^{+}, \quad V_{j} \sim \text{Beta}(1,\alpha) \text{ for } j \in \mathbb{Z}^{+},
     \label{eq:stick-breaking}
\end{split}
\end{equation}
where $\delta_{\cdot}$ is a measure with a point mass of 1 at $\cdot$. This representation makes clear that the distribution $P$ is discrete almost surely. The support of $P$ consists of a countably infinite set of possible values $\{\boldsymbol{\theta}_{1}, \boldsymbol{\theta}_{2},\dots\}$ drawn from the base distribution $P_{0}$, and the weights associated with these values are $\{w_{1},w_{2},\dots \}$. The term stick-breaking comes from an interpretation of the construction of $\{w_{1},w_{2},\dots\}$ as consecutively breaking a stick of length 1 into an infinite number of pieces. At the $j$-th step, the length of the remaining stick is $\prod_{l=1}^{j-1} (1-V_{l})$, and a piece of this stick is broken off according to the random variable $V_{j} \sim \text{Beta}(1,\alpha)$. The length of the stick we break off at the $j$-th step is assigned to $w_{j}$ such that $\sum_{j=1}^{\infty} w_{j} = 1$ with probability one. The scale parameter $\alpha$ influences the number of clusters produced by the stick-breaking construction, with larger $\alpha$ leading to larger number of clusters on average. We place a conjugate prior on $\alpha$ as $\alpha \sim \text{Gamma}(a_0,b_0)$ as is commonly used for stick-breaking Dirichlet process priors \citep{Hastieetal2015,IshwaranJames2001,BleiJordan2006}. 

The prior on the cluster index variable $Z_{m}$, $m=1,\dots,M$, has an explicit form. The index $Z_m$ takes value $j \in \mathbb{Z}^{+}$ with probability $w_{j}$: 
\begin{align*}
    &\Pr(Z_{m} = j) = w_{j} \quad \text{for} ~j \in \mathbb{Z}^{+}, \quad m = 1,\dots,M.
\end{align*}
The stick-breaking construction naturally leads to an ordering of the cluster labels based on the sizes of the clusters, due to the fact that $\mathbb{E}( w_{j} ) > \mathbb{E} ( w_{j+1} )$ for all $j\in \mathbb{Z}^{+}$ \citep{PapaspiliopoulosRoberts2008}. Throughout this paper, we assume that the cluster labels are arranged in order of decreasing cluster size.

We complete the model formulation by specifying priors for the regression parameters and covariance parameters. We place independent priors on the regression parameters, given by $\boldsymbol{\beta}_{q} \sim \mathcal{N}_M(0, \tau^{2}\boldsymbol{I}_{M})$, $q=1,\dots,p$, where $\tau$ is a hyperparameter to be specified and $\boldsymbol{I}_{M} \in \mathbb{R}^{M\times M}$ is the identity matrix. We do not assume an underlying group structure for the regression coefficients $\boldsymbol{\beta}_{q}$. This modeling choice is due to the nature of our motivating ABIDE data in Section \ref{sec:data}, where the covariates are measured at baseline for each individual and do not vary over $M$. We specify the prior $\sigma_{j}^2 \sim \text{Inv-Gamma}(a_1,b_1)$ independently for the variance parameters, $j=1,2, \ldots$ . We specify the prior distribution $\eta_{j} \sim \text{Beta}(a_2,b_2)$ for the correlation parameters, where $\eta_{j}$ is a suitable transformation of $\rho_j$ into $(0,1)$ based on the correlation function $\boldsymbol{\Gamma}(\cdot, \cdot; \rho_j)$, with examples given in Section \ref{sec:sim}. 

\subsection{The complete hierarchical model formulation}

To avoid the challenging sampling of the infinite-dimensional $\{V_{1},V_{2},\dots\}$ in equation \eqref{eq:stick-breaking}, we propose to use the truncated Dirichlet process introduced by \cite{IshwaranJames2001}. In this approach, we set $V_{K}$ to 1, where $K$ is an \emph{a priori} fixed value corresponding to the maximum number of clusters allowed in a partition of $\mathcal{S}$. By doing so, we ensure that for a sufficiently large $K$, the infinite sum of stick-lengths, $\sum_{j=1}^{\infty} w_{j} = 1$, can be transformed into a finite sum, such that $\sum_{j=1}^{K} w_{j} = 1$ almost surely. This truncation allows the truncated Dirichlet process to closely approximate the Dirichlet process. We denote the vector of finite-dimensional stick end-points as $\boldsymbol{V} = (V_{1},\dots,V_{K})$. By default, we set $K=M$. In situations where $M$ is sufficiently large or when prior knowledge from subject matter science suggests a smaller number of clusters, $K$ can be reduced, e.g. $K=M/2$. Conversely, when $M$ is small, we set $K$ to be large, e.g. $K=2M$, to minimize the error introduced by the truncation. 

The truncation $\boldsymbol{V}$ leads to the truncation of the number of covariance parameters at $K$. We therefore define $\boldsymbol{\sigma}^2 = (\sigma_{1}^2, \dots,\sigma_{J}^2,\sigma_{J+1}^2,\dots, \sigma_{K}^2)$ and $\boldsymbol{\rho} = (\rho_{1}, \dots,\rho_{J},\rho_{J+1},\dots, \rho_{K})$. The complete hierarchical model formulation is given by:
\begin{align}
    &\alpha \sim \text{Gamma}(a_0,b_0) \nonumber \\ 
    &\boldsymbol{\beta}_{q} \sim \mathcal{N}_M(0, \tau^{2} \boldsymbol{I}_{M}) \text{ for } q=1,\dots,p \nonumber \\ 
    &\boldsymbol{\theta}_{j} = (\sigma_{j}^{2}, \eta_{j}) \sim \text{Inv-Gamma}(a_1,b_1) \cdot \text{Beta}(a_2, b_2) \text{ for } j = 1,\dots,K \nonumber\\
    &V_{j} \,|\, \alpha \sim \text{Beta}(1,\alpha) \text{ for } j = 1,\dots,K-1 \text{ and } V_{K} = 1 \nonumber \\ 
    &w_{j} = V_{j} \prod_{l=1}^{j-1} (1-V_{l}) \text{ for } j = 1,\dots,K \nonumber \\ 
    &\boldsymbol{Z}\,|\,\boldsymbol{V} \,\sim \text{Categorical}_{K}\{ w(\boldsymbol{V}) \} \nonumber \\ 
    &\boldsymbol{\Sigma}_{\text{perm}}(\mathcal{S},\mathcal{
    S}; \boldsymbol{\theta}) = \text{Block-diag}\{ \boldsymbol{\Sigma}_{\text{perm}}(\mathcal{S}_{1},\mathcal{
    S}_{1}; \boldsymbol{\theta}_{1}), \dots, \boldsymbol{\Sigma}_{\text{perm}}(\mathcal{S}_{J},\mathcal{
    S}_{J}; \boldsymbol{\theta}_{J}) \} \nonumber \\
    &\boldsymbol{\Sigma}(\mathcal{S}, \mathcal{S}; \boldsymbol{\theta}) = \boldsymbol{P}_{\pi}^{\top} \boldsymbol{\Sigma}_{\text{perm}}(\mathcal{S}, \mathcal{S}; \boldsymbol{\theta}) \boldsymbol{P}_{\pi} \nonumber \\
    &\boldsymbol{y}_i(\mathcal{S}) \,|\, \boldsymbol{B},\boldsymbol{Z}, \boldsymbol{\theta} \,\sim \mathcal{N}_M \bigl\{\boldsymbol{B}\boldsymbol{X}_{i}, \boldsymbol{\Sigma}(\mathcal{S}, \mathcal{S}; \boldsymbol{\theta}) \bigr\} \text{ for } i=1,\dots,N.  \nonumber 
\end{align}
The Gibbs sampler is used to draw samples from the conditional distributions of $\boldsymbol{V}, \boldsymbol{\sigma}^2, \alpha, \boldsymbol{B}$, while the slice sampler of \cite{LiWalker2020} is used for the independent sampling of $\eta_{j}$. The step size of the slice sampler is determined by the variance tuning parameter $\lambda$, with a larger $\lambda$ resulting in a larger step size and a smaller $\lambda$ leading to a smaller step size. We will return to the choice of $\lambda$ in Section \ref{sec:samplerZ} below, where we describe the steps involved in drawing samples of $\boldsymbol{Z}$. A comprehensive explanation of the process of sampling from the conditional distributions of $\boldsymbol{V}, \boldsymbol{\rho}, \boldsymbol{\sigma}^2, \alpha$, and $\boldsymbol{B}$ using a slice-within-Gibbs sampler is given in Section 1 of the Supplementary Material.


\section{Split-merge sampler for cluster indices \texorpdfstring{$\boldsymbol{Z}$}{Lg}} \label{sec:samplerZ}

\subsection{Poor mixing of Gibbs and slice samplers for \texorpdfstring{$\boldsymbol{Z}$}{Lg}} \label{sec:mixingZ}

Let $\boldsymbol{Y}(\mathcal{S}) = \{\boldsymbol{y}_{1}(\mathcal{S}),\dots,\boldsymbol{y}_{N}(\mathcal{S})\} \in \mathbb{R}^{M \times N}$ be the matrix of outcomes, and $\boldsymbol{Z}_{-m}$ represent the sub-vector of $\boldsymbol{Z}$ obtained by removing the $m$-th coordinate. We first consider a Gibbs sampler that draws each $Z_{m}$, $m \in \{1,\dots,M\}$, from its conditional posterior distribution, given by 
\begin{align*}
    p(Z_{m} = j \,|\, \boldsymbol{Z}_{-m}, \cdot) &\propto  p \bigl\{\boldsymbol{Y}(\mathcal{S}) \,\big|\, \boldsymbol{B}, \left( Z_{1}, \dots, Z_{m-1},j, Z_{m+1},\dots, Z_{M} \right), \boldsymbol{\theta} \bigr\} \nonumber \\
    &\quad \,\,  \cdot p\bigl\{ \left( Z_{1}, \dots, Z_{m-1},j, Z_{m+1},\dots, Z_{M} \right) \,\big|\, \boldsymbol{V}\bigr\}, \quad j=1, \ldots, K.
\end{align*}
Sampling the cluster indices $\boldsymbol{Z}$ in this fashion results in $O(MK)$ operations at each MCMC iteration, which is computationally intractable when $M$ and/or $K$ is large. The difficulty is exacerbated when the number of clusters $J$ is large, such that a large number of covariance parameters need to be estimated. Further, a large number of MCMC iterations are required to explore the space of partitions of $\mathcal{S}$ due to its size, which corresponds to the Bell \citep{bell1934exponential} number. Indeed, even small values of $M$, such as $M=10$, result in a Bell number that exceeds $10^5$. The difficulty in exploring this space efficiently causes many MCMC samplers for models that use the Dirichlet process prior to suffer from poor mixing issues even when the values of $M$ and/or $J$ are relatively small \citep{Hastieetal2015}. 

Walker's slice sampler \citep{Walker2014} offers an alternative solution for drawing samples of $Z_m$. The sampling steps for $Z_m$ in Walker's slice sampler are presented in Algorithm \ref{alg:SSZ}. At each MCMC iteration, Walker's slice sampler searches over the cluster indices $\{1, \ldots, \min(Z_{m}+s-1, K)\}$ for each outcome given some step size $s \in \mathbb{Z}^{+}$, which requires fewer operations on average compared to the $O(MK)$ complexity of the Gibbs sampler. By default, we set $s = K/2$.

\begin{algorithm}
\caption{Walker's slice sampler \citep{Walker2014} for $Z_{m}$}\label{alg:SSZ}
\begin{algorithmic}[1]

\REQUIRE Current values of $\boldsymbol{Z}$ and $z_{0}:= Z_{m}$; support of $Z_{m} \in \{1,\dots,K\}$; step size $s \in \mathbb{R}^{+}$
\ENSURE Proposed value of $Z_{m} := z_{1}$
\STATE Define $p_{0 | \boldsymbol{Z}_{-m}, \cdot}=p(Z_{m} = z_{0} \,|\, \boldsymbol{Z}_{-m},\cdot)$. Sample the slice variable $\omega \sim U(0, p_{0 | \boldsymbol{Z}_{-m}, \cdot})$, and sample 
\begin{align*}
    \ell_{1} \sim U\bigl\{z_{0}, \min(z_{0}+s-1, K)\bigr\}.
\end{align*} 

\STATE Set $a = 1$ and $b = \ell_{1}$.
\STATE Sample $Z^{*} \sim U\{a,b\}$. If $p(Z_{m} = z^{*} \,|\, \boldsymbol{Z}_{-m}, \cdot) > \omega$, accept $z_{1} = z^{*}$. Else, 
\begin{align*}
    \text{if } z^{*} < z_{0}  \text{ then } a = \max(a, z^{*}) \text{ else } b = \min(b,z^{*}).
\end{align*}
\STATE Repeat Step 3 until $p(Z_{m} = z^{*} \,|\, \boldsymbol{Z}_{-m}, \cdot) > \omega$ and set $z_{1} = z^{*}$.
\end{algorithmic}
\end{algorithm}

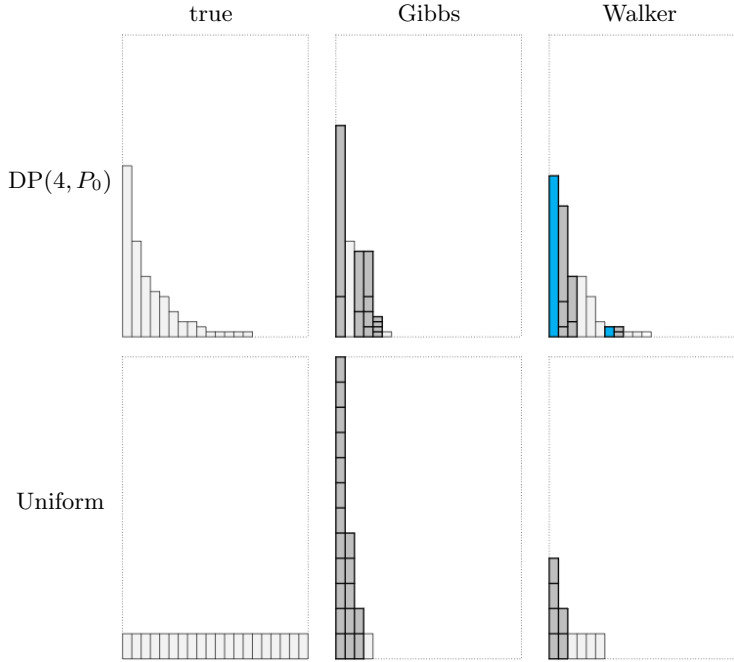
\begin{figure}[ht!]
\centering
\begin{tikzpicture}
    \node (image) at (0,0) {
        \includegraphics[width=0.5\textwidth]{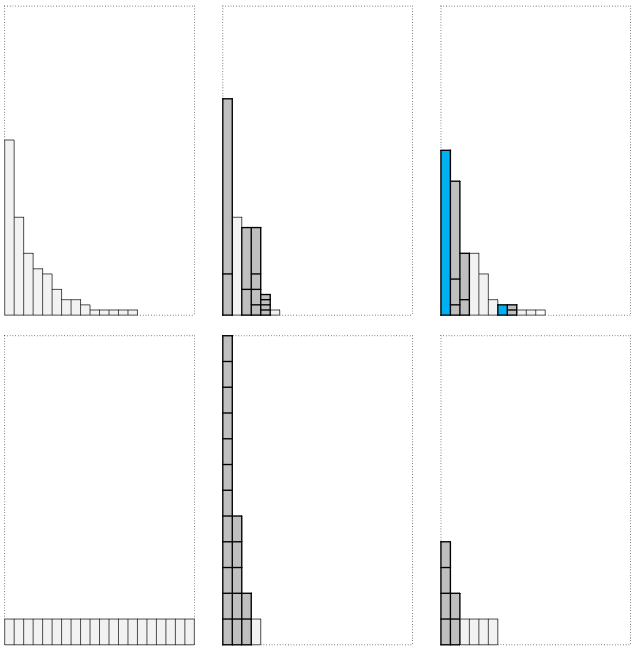}
    };
    \node[above,black,fill=white] at (-4.9,1.9){\small $\text{DP}(4,P_{0})$};
    \node[above,black,fill=white] at (-4.9,-2.3){\small Uniform};
    \node[above,black,fill=white] at (-2.9,4.2){\small true};
    \node[above,black,fill=white] at (0,4.2){\small Gibbs};
    \node[above,black,fill=white] at (2.8,4.2){\small Walker};
\end{tikzpicture}
\caption{The true partition (left) and MAP estimate of the partition obtained from the Gibbs sampler (middle) and Walker's slice sampler (right). The true partition is generated from a $\text{DP}(4,P_{0})$ (top) or a uniform distribution with equal partition sizes (bottom), with $M=100$ and $N=100$, assuming a generalized AR(1) covariance structure in each cluster. Highlighted grey bars indicate additional split steps needed to reach the true partition; highlighted blue bars indicate additional merge steps needed to reach the true partition. The MAP estimate of the partition obtained from our proposed clustering method recovers the true partition (see results in Section \ref{sec:sim}).}
\label{fig:sm_gibbs_ss}
\end{figure}

Figure \ref{fig:sm_gibbs_ss} illustrates the poor mixing behavior of the commonly used Gibbs sampler and Walker's slice sampler with an example. It depicts the maximum a posteriori (MAP) estimates of the conditional posterior distribution of $\boldsymbol{Z}$ obtained using the Gibbs sampler and Walker's slice sampler. The true partition is generated either from a $\text{DP}(4,P_{0})$ distribution ($J=14$) or from a uniform distribution with equal partition sizes ($J=20$). The data dimensions are $M=100$ and $N=100$, with a generalized AR(1) covariance structure in each cluster --  the specific correlation function and covariance parameters are given in Section \ref{sec:sim} below. Both the Gibbs sampler and Walker's slice sampler have difficulty escaping the local mode of the partition space, resulting in MAP estimates that fail to accurately identify the true partition.

In this example, however, Walker's slice sampler delivers a MAP that is closer to the true partition: it requires fewer split-merge steps to recover the true partition compared to the Gibbs sampler. Split steps involve selecting a cluster and dividing its elements into two clusters, while merge steps involve selecting two clusters and merging them into a single cluster. From Figure \ref{fig:sm_gibbs_ss}, we see that i) a larger number of split steps are typically needed to recover the true partition compared to merge steps, and ii) split steps are needed more frequently in clusters of larger and smaller sizes. 

These two observations motivate the development of a new sampling algorithm that improves the mixing of Gibbs and Walker's slice samplers for $\boldsymbol{Z}$ by performing split-merge steps in addition to Walker's slice sampler. The Metropolis-Hastings rejection-sampling approach for split-merge steps has been used with standard Gibbs steps to address the issue of slow convergence and poor mixing of the Gibbs sampler \citep{GreenRichardson2001, Dahl2003, JainNeal2004, WangRussell2015} for models using the Dirichlet process prior, particularly in the context of Dirichlet process mixture models. Although these existing split-merge samplers have proven effective, they are primarily tailored for settings with a small $M$ and small number of clusters $J$. In contrast, our focus is on the setting where both $M$ and $J$ are large, potentially with a greater number of small clusters, including single-outcome clusters. To address this high-dimensional setting, we propose the Head-Tail Split-Merge (HTSM) sampler in Section \ref{sec:HTSM} below.

\subsection{Head-Tail Split-Merge (HTSM) sampler} \label{sec:HTSM}

We define the head of the partition as the larger clusters, and the tail as the smaller clusters. Our proposed HTSM sampler offers two types of split-merge moves: head or tail. A head split-merge is a split or merge move applied to the head of a partition, and a tail split-merge is a split or merge move applied to the tail of a partition. The HTSM sampler is structured into three phases, illustrated in Figure \ref{f:HTSM-phases}. The first and second phases serve as the burn-in period of the MCMC, aimed at reaching convergence of the sampler. The third phase corresponds to the sampling period of the MCMC, where samples are generated for analysis and inference. Specifically:

\begin{figure}[ht!]
\centering
\begin{tikzpicture}
    \draw[thick,<->] (0,0) -- (2,0);
    \draw[thick,<->] (2,0) -- (4,0);
    \draw[thick,<->] (4,0) -- (10,0);

    \node[above,fill=white] at (1,0.1){\small phase I};
    \node[above,fill=white] at (3,0.1){\small phase II};
    \node[above,fill=white] at (7,0.1){\small phase III};
    \node[above,fill=white] at (2,0.5){\small [burn-in period]};
    \node[above,fill=white] at (7,0.5){\small [sampling period]};
\end{tikzpicture}
    \caption{Phases of the proposed HTSM sampler. \label{f:HTSM-phases}}
\end{figure}
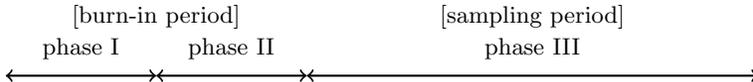

\begin{itemize}[itemsep=-0.5em, leftmargin=*]
\item \textbf{In phase I of the burn-in period}, we alternate between Walker's slice sampler and head split-merge moves proposed by our HTSM sampler. The HTSM sampler proposes a head split or merge with probabilities $p_0$ and $1-p_0$, respectively. These moves aim to detect larger clusters. 

\item \textbf{In phase II of the burn-in period}, the HTSM sampler proposes tail split-merge moves, where a tail split or merge is proposed with a probability of $p_0$ or $1-p_0$, respectively. These moves identify small clusters that may need to be split into smaller clusters.

\item \textbf{In phase III, the sampling period}, the HTSM sampler proposes tail split-merge moves, where a tail split or merge is proposed with a probability of $p_0$ or $1-p_0$, respectively. Further, we increase the variance tuning parameter $\lambda$ when sampling $\rho_j$ using the slice sampler of \citet{LiWalker2020}. This adjustment is motivated by \cite{Hastieetal2015}, who observed that samplers become trapped at a partition with high posterior support when insufficient variance is introduced in the sampling. Increasing the variance of the tuning parameter encourages the sampler to explore a broader range of potential partitions by introducing greater variability in $\rho_j$ and, subsequently, in $\boldsymbol{Z}$.
\end{itemize}

In each iteration of the MCMC, in addition to employing the HTSM sampler for $\boldsymbol{Z}$, we use the slice-within-Gibbs sampler as outlined in Section 1 of the Supplementary Material to sample the remaining parameters. Following the completion of the HTSM sampler in each iteration, we conduct a relabeling of $\boldsymbol{Z}$. This relabeling step is crucial to ensure the cluster labels are sorted from the largest to the smallest cluster. 

The proposed HTSM sampler is outlined in Algorithm \ref{alg:HTSM}. Our approach uses $p_{0}$ to decide between a split or merge move, allowing us to control the proportion of split and merge moves. We set the default value of $p_0$ to $0.7$ to encourage split moves to occur more frequently than merge moves. Our approach of using $p_{0}$ differs from \cite{JainNeal2004} and \cite{Dahl2003}, where a pair of indices are uniformly chosen to determine a split or merge move based on whether the selected pair is in the same cluster or not. \cite{WangRussell2015} used a similar approach to ours by pre-selecting the split-merge move type, but they chose their split-merge moves uniformly. As we show in Section \ref{sec:sim}, these differences have substantial benefits for the convergence of our sampler.

On the one hand, the split step across the three phases is characterized by two different vectors that encourage splitting in the large versus small clusters, respectively. On the other hand, merge steps across both phases give equal probability to all clusters, regardless of their size, to be merged. Below, we describe the probability vectors that favor large and small clusters, as well as the implementation details for the split-merge moves. Let $\boldsymbol{Z}^{*}$ and $\boldsymbol{\theta}^{*}$ be the candidate states which we will consider for the split-merge proposal. Denote $J^{+}$ as the number of clusters of size greater than one.

\begin{algorithm}[ht!]
\caption{Head-Tail Split-Merge sampler}\label{alg:HTSM}
\begin{algorithmic}[1]

\REQUIRE Current values of $\boldsymbol{Z}, \boldsymbol{\theta}$ ; split-merge  probability $p_{0}$; $\text{type}_{HT} \in \{\text{Head}, \text{Tail}\}$
\ENSURE Proposed values of $\boldsymbol{Z}, \boldsymbol{\theta}$
\STATE Set candidate states as $\boldsymbol{Z}^{*}$ and $\boldsymbol{\theta}^{*}$.
 
\STATE Choose the split-merge move type with probability 
\begin{align*}
    \Pr(\text{Split}) = p_{0} \text{ and } \Pr(\text{Merge}) = 1-p_{0}.
\end{align*}

\STATE If $\text{Split}$: \\
\quad If $\text{type}_{HT} = \text{Head}$, then $\boldsymbol{p}_{\text{split}} \in \mathbb{R}^{J^{+}}$ is a decreasing probability vector.\\
\quad Else if $\text{type}_{HT} = \text{Tail}$, then $\boldsymbol{p}_{\text{split}} \in \mathbb{R}^{J^{+}}$ is an 
 increasing probability vector.\\
\quad Choose the $j$-th cluster to split according $\boldsymbol{p}_{\text{split}}$.\\
\quad Prepare the launch state by randomly assigning indices $\{Z_{m}^{*}\}=\{Z_m^{*}: Z_m^{*}=j\}$ into one of the two clusters: $j$ or $(J+1)$. \\
\quad Set $\theta_{J+1}^{*} := \theta_{j}^{*}$.\\
\quad Assign $\{Z_{m}^{*}\}$ sequentially into the $j$ or $(J+1)$-th cluster according to probabilities 
\begin{align*}
    p(Z_{m}^{*} = j \,|\, \boldsymbol{Z}^{*}_{-m}, \boldsymbol{\theta}^{*}, \cdot) \text{ and } p(Z_{m}^{*} = J+1 \,|\, \boldsymbol{Z}^{*}_{-m}, \boldsymbol{\theta}^{*}, \cdot).
\end{align*}\\

\STATE Else if $\text{Merge}$:\\

\quad Choose the $j$ and $j^{'}$-th clusters to merge uniformly from the $J$ clusters.\\
\quad Assign the indices $\{Z_{m}^{*}: Z_m^{*}=j\}$ and $\{Z_{m}^{*}: Z_m^{*}=j'\}$ to the $j$-th cluster.\\
\quad Set $\theta_{j^{'}}^{*} := \theta_{j}^{*}$.\\

\STATE Calculate the Metropolis-Hastings acceptance probability 
\begin{align*}
    \alpha^{*} = \min \Biggl[1, \frac{q\{(\boldsymbol{Z},\boldsymbol{\theta}) \,|\, (\boldsymbol{Z}^{*},\boldsymbol{\theta}^{*}),\cdot \}}{q\{(\boldsymbol{Z}^{*},\boldsymbol{\theta}^{*}) \,|\, (\boldsymbol{Z},\boldsymbol{\theta}),\cdot\}}
    \frac{p\{\boldsymbol{Y}(\mathcal{S}) \,|\, \boldsymbol{Z}^{*}, \boldsymbol{\theta}^{*}, \cdot\}}{p\{\boldsymbol{Y}(\mathcal{S}) \,|\, \boldsymbol{Z}, \boldsymbol{\theta}, \cdot\}} \frac{p(\boldsymbol{Z}^{*}\,|\, \boldsymbol{V})}{p(\boldsymbol{Z}\,|\, \boldsymbol{V})} \frac{p(\boldsymbol{\boldsymbol{\theta}^{*}})}{p(\boldsymbol{\boldsymbol{\theta}})} \Biggr] ,
\end{align*}
and accept the proposal with probability $\alpha^{*}$.
\end{algorithmic}
\end{algorithm}

In the event of a head or tail merge proposal, where $(\boldsymbol{Z}^{*},\boldsymbol{\theta}^{*}) =(\boldsymbol{Z}^{\text{merge}},\boldsymbol{\theta}^{\text{merge}})$, the sampler selects two clusters, the $j$-th and the $j'$-th cluster uniformly at random from the set of $J$ clusters. The indices $\{Z_{m}^{*}: Z_m^{*}=j\}$ and $\{Z_{m}^{*}: Z_m^{*}=j'\}$ within both clusters are then assigned to the $j$-th cluster, with the covariance parameter $\boldsymbol{\theta}_{j'}^{*}$ updated by setting it equal to $\boldsymbol{\theta}_{j}^{*}$.

In the case of a head or tail split proposal, where $(\boldsymbol{Z}^{*},\boldsymbol{\theta}^{*}) =(\boldsymbol{Z}^{\text{split}},\boldsymbol{\theta}^{\text{split}})$, the initial step involves constructing a probability vector $\boldsymbol{p}_{\text{split}} \in \mathbb{R}^{J^{+}}$. On the one hand, if a head split is proposed, a decreasing probability vector $\boldsymbol{p}_{\text{split}} \in \mathbb{R}^{J^{+}}$ is constructed, with a preference for selecting larger clusters for the split move. On the other hand, if a tail split is proposed, an increasing probability vector $\boldsymbol{p}_{\text{split}} \in \mathbb{R}^{J^{+}}$ is constructed, and a cluster to split is randomly chosen based on this vector, resulting in a tendency to choose smaller clusters.

After selecting the $j$-th cluster for splitting according to $\boldsymbol{p}_{\text{split}}$, we prepare the launch state by randomly assigning the indices $\{Z_m^{*}\} = \{Z_m^{*}: Z_m^{*}=j\}$ within that cluster to either the same $j$-th cluster or a new $(J+1)$-th cluster. Additionally, the covariance parameter $\theta_{J+1}^{*}$ is updated by assigning it the value of $\theta_{j}^{*}$. Then, the indices $\{Z_m^{*}\}$ are sequentially assigned to one of the two clusters according to the conditional posterior distributions  $p(Z_{m}^{*} = j \,|\, \boldsymbol{Z}^{*}_{-m}, \boldsymbol{\theta}^{*}, \cdot)$ and $p(Z_{m}^{*} = J+1 \,|\, \boldsymbol{Z}^{*}_{-m}, \boldsymbol{\theta}^{*}, \cdot)$ using the restricted Gibbs sampler.

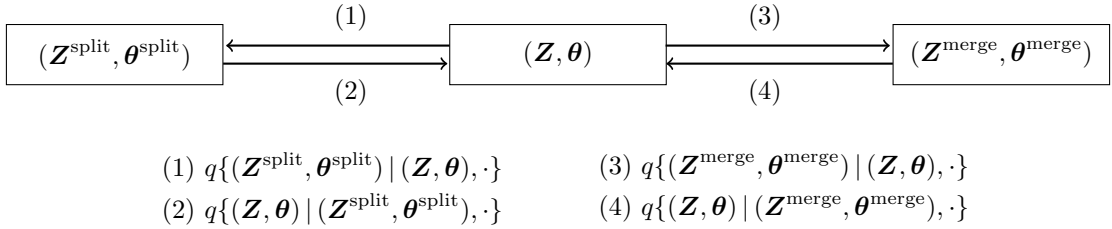
\begin{figure}[ht!]
\centering
\begin{tikzpicture}[
squarednode/.style={rectangle, draw=red!60, fill=red!5, very thick, minimum size=5mm},
mynode/.style={draw,fill=white,rectangle,minimum width=width("Magnetometerssss")+3pt,minimum height=0.8cm},
roundnode/.style={circle, draw=white, fill=white, minimum size=5mm,font=\small}
]
    \node[mynode]      (current)                        {$(\boldsymbol{Z}, \boldsymbol{\theta})$};
    \node[mynode]      (split)       [left=3 of current]  {$(\boldsymbol{Z}^{\text{split}}, \boldsymbol{\theta}^{\text{split}})$};
    \node[mynode]      (merge)       [right=3 of current] {$(\boldsymbol{Z}^{\text{merge}}, \boldsymbol{\theta}^{\text{merge}})$};

    \draw[thick,->] (-4.45,-0.12) -- (-1.47,-0.12);
    \draw[thick,<-] (-4.41,0.12) -- (-1.44,0.12);
    \draw[thick,->] (1.43,0.12) -- (4.41,0.12);
    \draw[thick,<-] (1.44,-0.12) -- (4.45,-0.12);

    \node[circle] at (-2.75,0.5){$(1)$};
    \node[circle] at (-2.75,-0.5){$(2)$};
    \node[circle] at (2.75,0.5){$(3)$};
    \node[circle] at (2.75,-0.5){$(4)$};

    \node[rectangle,fill=none] at (-3,-1.5){(1) $q\{(\boldsymbol{Z}^{\text{split}},\boldsymbol{\theta}^{\text{split}}) \,|\, (\boldsymbol{Z},\boldsymbol{\theta}),\cdot \}$};

    \node[rectangle,fill=none] at (-3,-2.05){(2) $q\{(\boldsymbol{Z},\boldsymbol{\theta}) \,|\, (\boldsymbol{Z}^{\text{split}},\boldsymbol{\theta}^{\text{split}}),\cdot \}$};

    \node[rectangle,fill=none] at (3,-1.5){(3) $q\{(\boldsymbol{Z}^{\text{merge}},\boldsymbol{\theta}^{\text{merge}}) \,|\, (\boldsymbol{Z},\boldsymbol{\theta}), \cdot\}$};

    \node[rectangle,fill=none] at (3,-2.05){(4) $q\{(\boldsymbol{Z},\boldsymbol{\theta}) \,|\, (\boldsymbol{Z}^{\text{merge}},\boldsymbol{\theta}^{\text{merge}}),\cdot \}$};

\end{tikzpicture}
\caption{The DAG presenting the transition probabilities for a split-merge proposal of the HTSM.}
\label{fig:DAG}
\end{figure}

It remains to calculate the Metropolis-Hastings acceptance probability for the HTSM sampler, which takes the form 
\begin{align}
    \alpha^{*} = \min \Biggl[1, \frac{q\{(\boldsymbol{Z},\boldsymbol{\theta}) \,|\, (\boldsymbol{Z}^{*},\boldsymbol{\theta}^{*}),\cdot \}}{q\{(\boldsymbol{Z}^{*},\boldsymbol{\theta}^{*}) \,|\, (\boldsymbol{Z},\boldsymbol{\theta}),\cdot\}}
    \frac{p\{\boldsymbol{Y}(\mathcal{S}) \,|\, \boldsymbol{Z}^{*}, \boldsymbol{\theta}^{*}, \cdot\}}{p\{\boldsymbol{Y}(\mathcal{S}) \,|\, \boldsymbol{Z}, \boldsymbol{\theta}, \cdot\}} \frac{p(\boldsymbol{Z}^{*}\,|\, \boldsymbol{V})}{p(\boldsymbol{Z}\,|\, \boldsymbol{V})} \frac{p(\boldsymbol{\boldsymbol{\theta}^{*}})}{p(\boldsymbol{\boldsymbol{\theta}})} \Biggr] , \label{eq:MHratio}
\end{align}
where $(\boldsymbol{Z}^{*},\boldsymbol{\theta}^{*})$ is either $(\boldsymbol{Z}^{\text{split}},\boldsymbol{\theta}^{\text{split}})$ or $(\boldsymbol{Z}^{\text{merge}},\boldsymbol{\theta}^{\text{merge}})$ depending on the type of proposal, and $q\{(\boldsymbol{Z},\boldsymbol{\theta}) \,|\, (\boldsymbol{Z}^{*}, \boldsymbol{\theta}^{*}),\cdot \}$ and $q\{(\boldsymbol{Z}^{*},\boldsymbol{\theta}^{*}) \,|\, (\boldsymbol{Z},\boldsymbol{\theta}),\cdot\}$ are transition probabilities from $(\boldsymbol{Z}^{*},\boldsymbol{\theta}^{*})$ to $(\boldsymbol{Z},\boldsymbol{\theta})$ and from $(\boldsymbol{Z},\boldsymbol{\theta})$ to $(\boldsymbol{Z}^{*},\boldsymbol{\theta}^{*})$ respectively. Here, we factored the posterior distribution into a product of the likelihood and the prior, and ignored terms not involving $(\boldsymbol{Z},\boldsymbol{\theta})$ or $(\boldsymbol{Z}^{*},\boldsymbol{\theta}^{*})$. 
A graphical presentation of the transition probabilities for HTSM split-merge proposals is shown as a directed acyclic graph (DAG) in Figure \ref{fig:DAG}. We can calculate the individual terms in Equation (\ref{eq:MHratio}) as follows. 

\begin{itemize}[itemsep=-0.5em, leftmargin=*]
\item The split proposal probability  $q\{(\boldsymbol{Z}^{\text{split}},\boldsymbol{\theta}^{\text{split}}) \,|\, (\boldsymbol{Z},\boldsymbol{\theta}),\cdot\}$ can be expressed as 
\begin{align*}
q\{(\boldsymbol{Z}^{\text{split}},\boldsymbol{\theta}^{\text{split}}) \,|\, (\boldsymbol{Z},\boldsymbol{\theta}),\cdot \}
&= q\{\boldsymbol{Z}^{\text{split}} \,|\, \boldsymbol{\theta}^{\text{split}}, (\boldsymbol{Z},\boldsymbol{\theta}),\cdot\} \,q\{\boldsymbol{\theta}^{\text{split}} \,|\, (\boldsymbol{Z},\boldsymbol{\theta}),\cdot\} \\
&= q(\boldsymbol{Z}^{\text{split}} \,|\, \boldsymbol{\theta}^{\text{split}}, \boldsymbol{Z},\cdot) \\
&= \Bigl(\frac{1}{2}\Bigr)^{d_{j}} \prod_{m \in \mathcal{I}_{d_{j}} } p(Z_{m} = z_{m}^{\text{split}} \,|\, \boldsymbol{Z}_{-m}, \boldsymbol{\theta}^{\text{split}}, \cdot),
\end{align*}
where $\mathcal{I}_{d_{j}}$ denotes the set of coordinates for the launch state, with a size of $d_{j}$. The second equality is derived by observing that splitting the cluster indices vector is only influenced by $(\boldsymbol{Z}, \boldsymbol{\theta}^{\text{split}})$, and not by $\boldsymbol{\theta}$. Additionally, $q\{\boldsymbol{\theta}^{\text{split}} \,|\, (\boldsymbol{Z},\boldsymbol{\theta}),\cdot\} = 1$ since, given $(\boldsymbol{Z}, \boldsymbol{\theta})$, there is only one way to obtain   $\boldsymbol{\theta}^{\text{split}}$, which is to set $\theta_{J+1}^{\text{split}} := \theta_{j}^{\text{split}}$. The last equality is the multiplication of two terms: the first term represents the probability of proposing a split move from the original state to the launch state (${1/2}^{d_{j}}$). The second term is the product of the sequential allocation proposal probabilities from the launch state to the split state. Specifically, each $p(Z_{m} = z_{m}^{\text{split}} \,|\, \boldsymbol{Z}_{-m}, \boldsymbol{\theta}^{\text{split}}, \cdot)$ denotes the probability that $Z_{m}$, the cluster index in the launch state, is allocated to the observed cluster index $z_{m}^{\text{split}}$ in the split state, where $z_{m}^{\text{split}}$ is either $j$ or $J+1$. The product is calculated sequentially for all $m \in \mathcal{I}_{d_{j}}$.

\item The reverse merge proposal probability is calculated as $q\{(\boldsymbol{Z},\boldsymbol{\theta}) \,|\, (\boldsymbol{Z}^{\text{split}},\boldsymbol{\theta}^{\text{split}}),\cdot \}= 1/J$. This probability is determined by uniformly selecting one cluster from a set of $J$ clusters for merging, while keeping the second cluster fixed as the $(J+1)$-th cluster. The two clusters are then merged into the first cluster that was chosen. 
\item The merge proposal probability is calculated as $q\{(\boldsymbol{Z}^{\text{merge}},\boldsymbol{\theta}^{\text{merge}}) \,|\, (\boldsymbol{Z},\boldsymbol{\theta}), \cdot\} = 1/\{J(J-1)\}$. The proposal uniformly selects two clusters from a set of $J$ clusters, and merges the two clusters into the first cluster that was chosen. 
\item The reverse split proposal probability $q\{(\boldsymbol{Z},\boldsymbol{\theta}) \,|\, (\boldsymbol{Z}^{\text{merge}},\boldsymbol{\theta}^{\text{merge}}), \cdot \}$ is calculated in the same way as the split proposal probability as:
\begin{align*}
q\{(\boldsymbol{Z},\boldsymbol{\theta}) \,|\, (\boldsymbol{Z}^{\text{merge}},\boldsymbol{\theta}^{\text{merge}}),\cdot \} 
&= q\{\boldsymbol{Z} \,|\, \boldsymbol{\theta}, (\boldsymbol{Z}^{\text{merge}},\boldsymbol{\theta}^{\text{merge}}),\cdot \}\,q\{\boldsymbol{\theta} \,|\, (\boldsymbol{Z}^{\text{merge}},\boldsymbol{\theta}^{\text{merge}}),\cdot\} \\
&= q(\boldsymbol{Z} \,|\, \boldsymbol{\theta}, \boldsymbol{Z}^{\text{merge}},\cdot) \\
&= \Bigl(\frac{1}{2}\Bigr)^{d_{j}} \prod_{m \in \mathcal{I}_{d_{j}} } p(Z^{\text{merge}}_{m} = z_{m} \,|\, \boldsymbol{Z}^{\text{merge}}_{-m}, \boldsymbol{\theta}, \cdot)
\end{align*}
where $\mathcal{I}_{d_{j}}$ denotes the set of coordinates for the launch state, with a size of $d_{j}$.
\end{itemize}

\subsection{Convergence of hybrid samplers}

Li and Walker's slice sampler \citep{LiWalker2020}, employed for sampling the correlation parameters $\eta_{j}$, is an extension of Neal's well-known slice sampler \citep{Neal2003}, featuring a modified search component. Additionally, Walker's slice sampler \cite{Walker2014}, used for sampling the cluster indices $\boldsymbol{Z}$, is an adaptation of Li and Walker's slice sampler tailored for discrete sample spaces.  The convergence properties of slice samplers have been thoroughly investigated. \cite{RobertsRosenthal1999} demonstrated the geometric ergodic properties of slice samplers, while \cite{MiraTierney2002} established a sufficient condition for the uniform ergodicity of slice samplers. Moreover, \cite{Walker2014} showed the convergence of their proposed transition density to the stationary distribution when the step size is large. The ergodicity of the HTSM sampler is ensured as it operates as a Metropolis-Hastings split-merge algorithm \citep{mengersen1996rates}. For an in-depth discussion on the ergodicity of Metropolis-Hastings split-merge algorithms, refer to Section 3.4 of \citet{Neal2003}. 

Our sampling approach involves a hybrid sampler, combining Walker's slice sampler, Li and Walker's slice sampler, Gibbs sampler, and the HTSM sampler. The convergence of hybrid samplers involving multiple distinct MCMC algorithms has been extensively investigated. As detailed in Section 2.4 of \citet{tierney1994markov}, combining Metropolis-Hastings and Gibbs sampling steps yields an ergodic Markov chain. Following the findings of \citet{MiraTierney2002}, the slice-within-Gibbs sampler also maintains ergodicity.

\section{Numerical illustrations using simulated data} \label{sec:sim}

Two sets of simulated data scenarios with large values of $M$ and $J$ are explored. To mimic our real data example, we set $N$ to be of the same order as $M$. In the first simulation, the true partition is generated from a Uniform distribution with equal cluster sizes. In the second simulation, the true partition is generated from a Dirichlet process prior. We consider three correlation functions $\boldsymbol{\Gamma}(\cdot, \cdot; \rho_j)$ at various points throughout the two simulations: Mat\'{e}rn, generalized AR(1) and compound symmetry, described below. In each simulation, results are compared across 50 simulated data sets each run for $10,000$ MCMC iterations. The burn-in period, comprising two phases, consists of $1,000$ iterations each, resulting in a total of $2,000$ iterations. The remaining $8,000$ iterations are allocated for the sampling period. Throughout the burn-in period, the variance tuning parameter $\lambda$ for $\rho_{j}$ is set to $100$, while during the sampling period, it is set to $150$.

In both simulations, $\boldsymbol{X}_{i}$ consists of $p=5$ covariates, including an intercept $X_{i1}$. The covariates $X_{i2}$ and $X_{i3}$ are generated from Bernoulli distributions with success probabilities of $0.5$ and $0.7$, respectively. The covariate $X_{i4}$ is generated from a standard Gaussian distribution, and the last covariate $X_{i5}$ is the interaction term between $X_{i2}$ and $X_{i4}$. 
We specify the maximum number of allowable clusters $K$ to be $M/2$. The prior specifications include setting $\tau = 1$ for the prior on regression coefficients, $\boldsymbol{\beta}_{q} \sim \mathcal{N}_M(\boldsymbol{0}, \boldsymbol{I}_{M})$, $q=1,\dots,p$. The variance parameters have prior distribution $\sigma_{j}^2 \sim \text{Inv-Gamma}(2.01,1.01)$, and the transformed correlation parameters have prior distribution $\eta_{j} \sim \text{Beta}(2.01,1.01)$. With the Mat\'ern and generalized AR(1) correlation functions, $\rho_j$ lies in $(0,1)$, so $\eta_j = \rho_j$. With the compound symmetry correlation function, $\rho_j \in (-1/(d_{j} - 1), 1)$, and its support depends on $\boldsymbol{Z}$. This poses challenges when updating $\rho_j$ during the MCMC since the support changes. To address this, we slightly narrow the support to $\rho_j \in (-1/(M - 1), 1)$, which ensures positive-definiteness of covariance matrices while maintaining a constant support throughout the iterations. The corresponding transformation is $\eta_{j} = (M-1)\rho_{j}/M + 1/M$. It is worth noting that the upper bound for the support of $\rho_{j}$ is typically 1 for most covariance structures. To avoid computational issues arising from near-singular correlation sub-matrices, we set the upper bound for $\rho_{j}$ to be $0.95$. The Dirichlet process scale parameter is assigned the prior $\alpha \sim \text{Gamma}(K + 0.01, 1.01)$. This choice ensures that the prior mean for $\alpha$ is approximately equal to $K$, effectively setting the initial number of clusters as large as possible. As discussed in \cite{Hastieetal2015}, it is recommended to set the initial number of clusters to be large in order to achieve convergence of the posterior distribution of $\alpha$.

For a head split proposal in the HTSM sampler, we set $\boldsymbol{p}_{\text{split}}$ such that the approximate probabilities of selecting clusters for splitting are as follows: about $0.3$ for the largest cluster, about $0.2$ for the second largest cluster, about $0.15$ for the third largest cluster, and about $0.1$ for the fourth largest cluster. This configuration results in around $75\%$ of the split moves occurring within these four largest clusters. Conversely, for a tail split proposal in the HTSM sampler, we configure $\boldsymbol{p}_{\text{split}}$ to be the reverse of that in the head split proposal.

In the first simulation, the performance of the model is evaluated when the true partition is from a Uniform distribution with equal cluster sizes, $J=M/5$. We consider three different settings. The first setting considers dimensions $M=500$ and $N=500$ with a homogeneous compound symmetric (CS) covariance structure, where the same $\rho$ and $\sigma^2$ are assumed for all clusters. The true $\rho$ and $\sigma^2$ are set to $0.7$ and $1.1$, respectively. The second setting also considers dimensions $M=500$ and $N=500$, but with a more complex heterogeneous generalized AR(1) covariance structure \citep{MurrayHelms1990}. Here, different sets of $\rho_{j}$ and $\sigma_{j}^2$ are assumed for each cluster. For each cluster, the true $\rho_{j}$ is sampled from a uniform sequence with a spacing of $0.05$ between $0.4$ and $0.9$, and the true $\sigma_{j}^2$ is sampled from a uniform sequence with a spacing of $0.05$ between $0.95$ and $1.5$. The generalized AR(1) covariance structure is defined by: 
\begin{align*}
    \text{Cov}\{y_{i}(\boldsymbol{s}_{\ell}), y_{i}(\boldsymbol{s}_{k})\} = 
    \sigma_{j}^2 \rho_{j}^{\{d(s_{\ell},s_{k})\}^{\nu}}\mathbbm{1}(\ell \ne k) + \sigma_{j}^2  \mathbbm{1}(\ell=k).
\end{align*}
Here, $\boldsymbol{s}_{m}\equiv s_m \in \mathbb{R}$, $m=1,\dots,M$ with $d(s_{\ell},s_{k})$ the Euclidean distance between two locations. The decay speed parameter is set to $\nu = 0.2$. The third and most complicated setting considers dimensions $M=1000$ and $N=1000$, and uses a heterogeneous Mat\'ern covariance structure \citep{Matern1986} with shape parameter $\nu = 1.5$, given by 
\begin{align*}
    \text{Cov}\{y_{i}(\boldsymbol{s}_{\ell}), y_{i}(\boldsymbol{s}_{k})\} = 
    \sigma_{j}^2 \Bigl\{1+\frac{\sqrt{3}d(s_{\ell},s_{k})}{\rho_{j}} \Bigr\} \exp\Bigl\{ -\frac{\sqrt{3}d(s_{\ell},s_{k})}{\rho_{j}} \Bigr\} \mathbbm{1}(\ell \ne k) + \sigma_{j}^2  \mathbbm{1}(\ell=k).
\end{align*}
Each $\rho_{j}$ and $\sigma_{j}^2$ is generated in the same manner as in the second setting. Here, $\boldsymbol{s}_{m} \in \mathbb{R}^{2}$, with $d(\boldsymbol{s}_{\ell},\boldsymbol{s}_{k})$ the Euclidean distance between two locations divided by 20. 

\begin{figure}[ht!]
\centering
\includegraphics[width=1\textwidth]{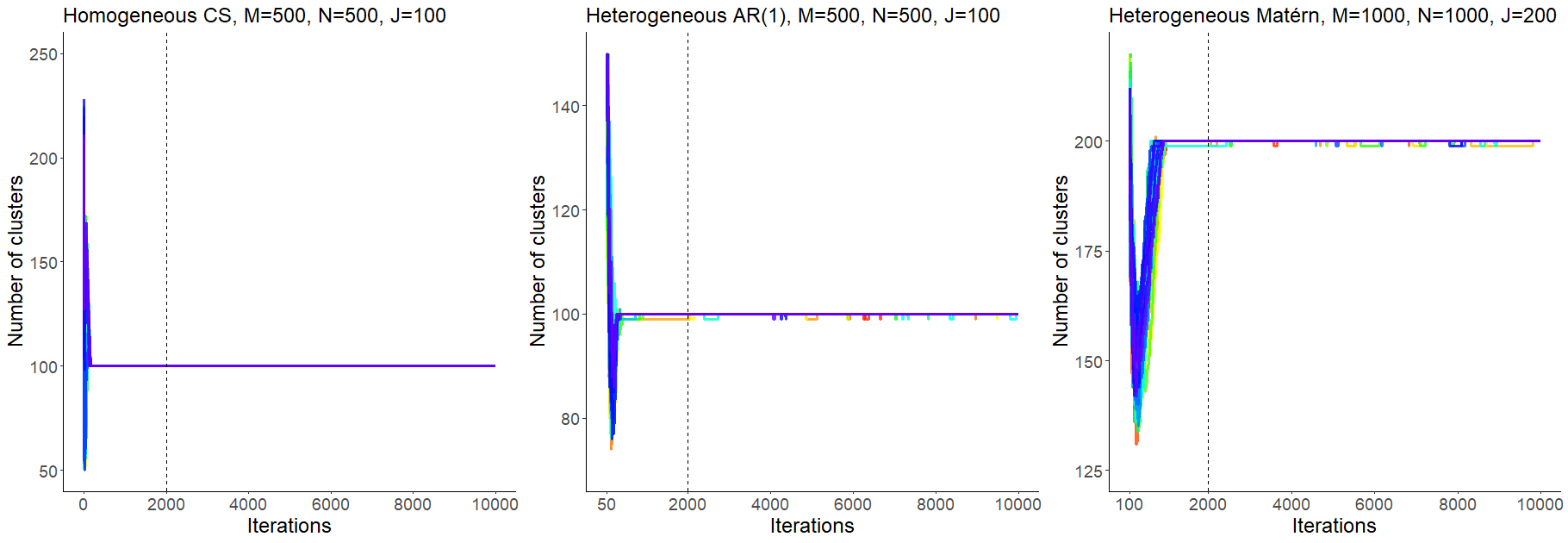}
\caption{First simulation -- Plot of the posterior of the number of clusters for the 50 simulations conducted over $10,000$ iterations, assuming a uniform true partition. Each line represents the posterior number of clusters for an individual simulation. The vertical dotted line at iteration $2,000$ indicates the end of the burn-in period.}
\label{fig:Jplot_uniform}
\end{figure}

Figure \ref{fig:Jplot_uniform} displays the posterior of the number of clusters across the three settings. As the complexity of the covariance structure increases, the convergence of the sampler requires a higher number of iterations. During the burn-in period, the MAP estimate in all 50 simulations of homogeneous CS correctly identified the true partition. For the heterogeneous AR(1) setting, the MAP estimate corresponded to the true partition in 49 out of 50 simulations, while in the heterogeneous Mat\'ern setting, it corresponded to the true partition in 48 out of 50 simulations. During the sampling period, for all three settings, the MAP estimate corresponded to the true partition in all 50 simulations.

Table \ref{table:CR_uniform} presents the average coverage rates for a 95\% credible interval across 50 simulations for $\boldsymbol{\rho}, \boldsymbol{\sigma}^2$ and $\boldsymbol{B}$ during the sampling period. The results indicate that the desired 95\% level is achieved for all three settings, with slightly conservative inference. In the homogeneous CS setting, the dimensions of $\rho$ and $\sigma^2$ are both one, making it relatively easy for the model to identify the distribution of these two parameters accurately. It should be noted that the coverage rates for $\boldsymbol{\rho}, \boldsymbol{\sigma}^2, \boldsymbol{B}$ are computed from the subset of iterations where $\boldsymbol{Z}$ is equal to its MAP estimate. This is because the posterior distributions for $\boldsymbol{\rho}, \boldsymbol{\sigma}^2$, and $\boldsymbol{B}$ depend on the value of $\boldsymbol{Z}$. When the estimate of $\boldsymbol{Z}$ is close to the true partition, the conditional posterior distributions are likely centered around the true values. Conversely, for estimates of $\boldsymbol{Z}$ that are farther from the true partition, the posterior distributions are likely centered around other values, leading to multi-modal marginal posterior distributions.

\begin{table}[ht!]
\caption{First simulation -- Coverage rate (standard deviation) of 95\% credible interval for each parameter, averaged across 50 simulations during the sampling period, assuming a uniform true partition. The coverage rates for $\boldsymbol{\rho}, \boldsymbol{\sigma}^2, \boldsymbol{B}$ are computed from the subset of iterations where $\boldsymbol{Z}$ is equal to its MAP estimate. }
\centering
\begin{tabular}{lcccccc}
& & & & \multicolumn{3}{c}{coverage rate (sd)}\\
\cline{5-7}
                      & $M$ & $N$ & $J$ & $\boldsymbol{\rho}$ & $\boldsymbol{\sigma}^2$ & $\boldsymbol{B}$ \\ \hline
homogeneous CS        & 500 & 500 & 100 & 1.00 (0.00) & 1.00 (0.00) & 0.95 (0.01)\\
heterogeneous AR(1)     & 500 & 500 & 100 & 0.97 (0.02) & 0.96 (0.02) & 0.95 (0.01)\\
heterogeneous Mat\'ern  & 1000 & 1000 & 200 & 0.98 (0.01) & 0.98 (0.01) & 0.95 (0.01)\\ \hline
\end{tabular}
\label{table:CR_uniform}
\end{table}

In the second simulation, we consider two settings with the true partition from a $\text{DP}(6,P_{0})$. The first setting has dimensions $M=500$, $N=500$, $J=23$, and a heterogeneous CS covariance structure. The second setting also has a heterogeneous CS covariance structure, but with larger dimensions $M=1000$, $N=1000$, and $J=37$. The sets of $\rho_{j}$ and $\sigma_{j}^2$ are sampled in the same way as in the first simulation. This setting allows for a small number of very large clusters and a moderate number of size one clusters in the true partition.

\begin{figure}[ht!]
\centering
\includegraphics[width=0.7\textwidth]{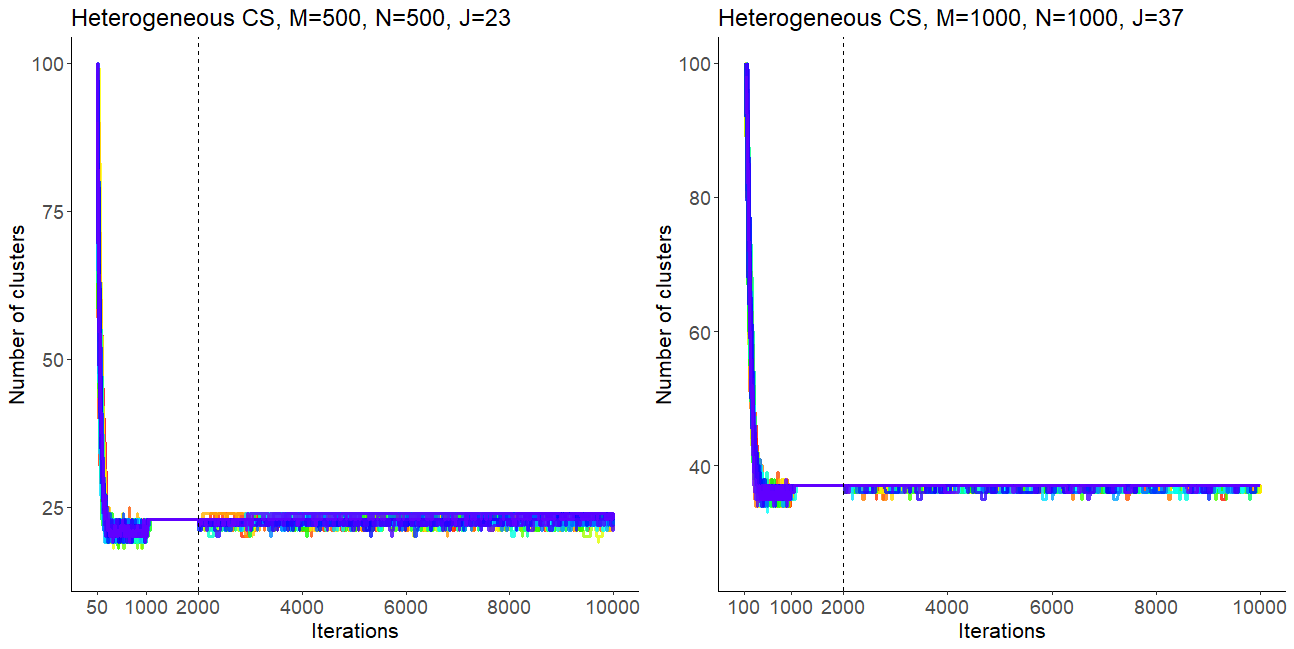}
\caption{Second simulation -- Plot of the posterior of the number of clusters for the 50 simulations conducted over $10,000$ iterations, assuming a $\text{DP}(6,P_{0})$ true partition. Each line represents the posterior number of clusters for an individual simulation. The vertical dotted line at iteration $2,000$ indicates the end of the burn-in period.}
\label{fig:Jplot_DP}
\end{figure}

Figure \ref{fig:Jplot_DP} illustrates the posterior distribution of the number of clusters across the two settings. Throughout the burn-in phase, the MAP estimate consistently identified the true partition in all 50 simulations in both settings. During the subsequent sampling phase, the MAP estimate corresponded to the true partition in all 50 simulations of the heterogeneous CS with $M=1000$ setting, and in 42 out of 50 simulations of the heterogeneous CS with $M=500$ setting. 

Table \ref{table:CR_DP} displays the average coverage rates of 95\% credible intervals across 50 simulations for $\boldsymbol{\rho}$, $\boldsymbol{\sigma}^2$, and $\boldsymbol{B}$ during the sampling period. The desired 95\% level of coverage was achieved for all parameters in both settings.

\begin{table}[ht!]
\caption{Second simulation -- Coverage rate (standard deviation) of 95\% credible interval for each parameter, averaged across 50 simulations during the sampling period, assuming a $\text{DP}(6,P_{0})$ true partition. The coverage rates are computed from the subset of iterations where $\boldsymbol{Z}$ is equal to its MAP estimate.
}
\centering
\begin{tabular}{lcccccc}
& & & & \multicolumn{3}{c}{coverage rate (sd)}\\
\cline{5-7}
                      & $M$ & $N$ & $J$ & $\boldsymbol{\rho}$ & $\boldsymbol{\sigma}^2$ & $\boldsymbol{B}$ \\ \hline
heterogeneous CS      & 500 & 500 & 23 & 0.97 (0.04) & 0.96 (0.04) & 0.95 (0.02)\\
heterogeneous CS      & 1000 & 1000 & 37 & 0.98 (0.02) & 0.97 (0.03) & 0.95 (0.02)\\ \hline
\end{tabular}
\label{table:CR_DP}
\end{table}

The increased variability during the sampling phase in both settings of the second simulation can be attributed to setting the variance tuning parameter of the slice sampler for $\rho_j$ to $\lambda = 150$. By increasing $\lambda$, the sampler exhibits more variability, allowing it to explore a wider range of potential partitions. However, this increased variability also means that the MAP estimate may not always match the true partition for some simulations. Nevertheless, it is important to note that the 95\% credible interval for $\boldsymbol{Z}$ across all 50 simulations contains the true partition in each setting. 

The second simulation only considers the CS covariance structure. When we consider covariance functions that rely on distances between measurements, estimating the true partition becomes challenging, particularly when dealing with large cluster sizes. In the case of a very large cluster, even with a slow decay rate, the off-diagonal terms of the covariance sub-matrix can be small, making it difficult to differentiate this covariance matrix from a combination of several moderately sized sub-matrices. It is important to highlight that this challenge primarily arises in high-dimensional settings where $M$ is large. Recalling the illustrative example in Figure \ref{fig:sm_gibbs_ss}, we were able to successfully identify the true partitions using a generalized AR(1) structure for a moderately sized problem ($M=100$ and $N=100$) in the first simulation.

\section{Estimating brain co-activation regions} \label{sec:data}

\subsection{ABIDE data and problem setup}

We use our proposed Bayesian clustering approach for dependence structures to identify regions of co-activation in the ABIDE dataset \citep{DiMartinoetal2014}. We consider rfMRI connectivity between $M=1000$ ROIs defined by the hierarchical multi-resolution parcellation introduced in \cite{Schaeferetal2018}. From the various versions of the Schaefer parcellation, we specifically adopt the 17-network parcellation provided by \cite{Kongetal2021}.
Our goal is to investigate between-network functional connectivity by identifying clusters of correlated ROIs that display co-activation patterns in school-age, male children diagnosed with both ASD and ADHD relative to those without co-occurring ADHD. To achieve this, we compare two distinct groups of individuals selected from the same participating sites within the dataset. The first group consists of 21 school-age, male children diagnosed with both ASD and ADHD, referred to as the ASD-coADHD group. The second group comprises 52 school-age, male children diagnosed with ASD but without ADHD, referred to as the ASD-nonADHD group. 

In each group, we denote the number of subjects as $n$. The length of the rfMRI time series with a lag of 2 for the $\ell$-th subject is represented by $T_{\ell}$. By incorporating a lag of 2, we ensure that the time points are sufficiently spaced apart and independent. The total length of the concatenated rfMRI outcomes for all $n$ individuals is defined as $N = \sum_{\ell=1}^{n} T_{\ell}$. Specifically, for the ASD-coADHD group, $N=1755$, and for the ASD-nonADHD group, $N=4594$. As the rfMRI outcome is an average across voxels within each ROI, the outcomes are approximately independently multivariate Gaussian due to the Central Limit Theorem. The covariates $\boldsymbol{X}_i \in \mathbb{R}^{n}$, $i = 1, \dots, N$, correspond to $\boldsymbol{e}_{\ell} \in \mathbb{R}^n$ for each $\ell \in \{1, \dots, n\}$. This representation gives a covariate corresponding to the indicator that the $i$-th outcome belongs to a time point for the $\ell$-th individual. 
The rfMRI time series $\boldsymbol{Y}(\mathcal{S})$ are centered and scaled. We assume the covariance of $\boldsymbol{Y}(\mathcal{S})$ is block-diagonal with blocks following the compound symmetric (CS) model. This choice is motivated by our interest in exploring functional connectivity between ROIs, which does not necessarily depend on spatial distance between the ROIs.

\subsection{Results from estimating functional connectivity}

For both the ASD-coADHD group and the ASD-nonADHD group separately, we perform MCMC runs for $14,000$ iterations, with a burn-in of $4,000$ iterations, for each of 50 different seeds. The MCMC tuning parameters and prior specifications are set to be consistent with those in Section \ref{sec:sim}. The results obtained from the 50 different seeds are combined to obtain the maximum a posteriori (MAP) estimate of $\boldsymbol{Z}$ and an estimate of the posterior mode for $\boldsymbol{\rho}$ conditioned on the MAP estimate of $\boldsymbol{Z}$. To evaluate the convergence of the sampler, we investigated traceplots and autocorrelation functions of the covariance parameters.

\begin{figure}[ht!]
\centering
\includegraphics[width=1\textwidth]{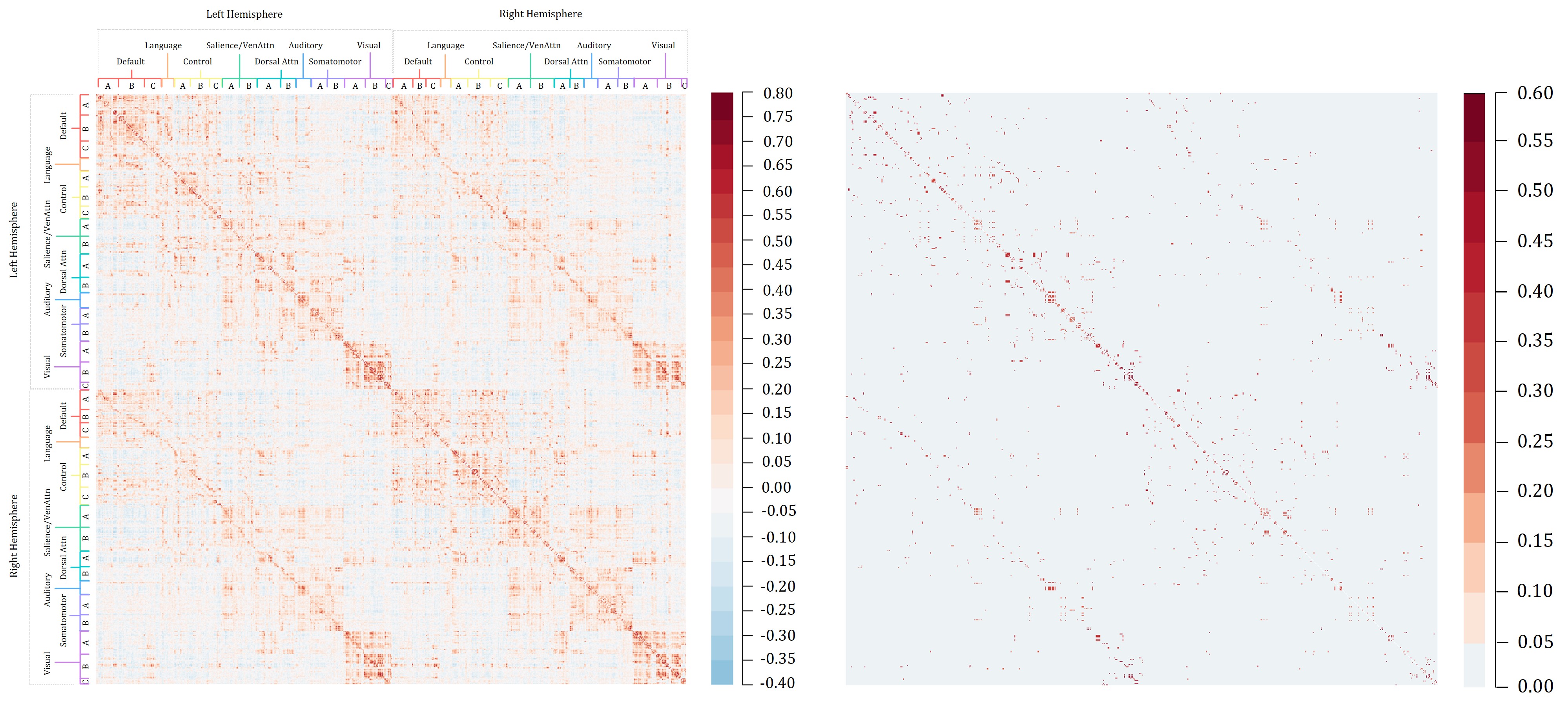}
\caption{Sample correlation matrix (left) and the estimated correlation matrix (right) for the ASD-coADHD group. The initial 500 ROIs correspond to the 17 networks of the left hemisphere, while the latter 500 ROIs correspond to the 17 networks of the right hemisphere.}
\label{fig:Cormat1000}
\end{figure}

Figure \ref{fig:Cormat1000} displays the sample correlation matrix and the estimated correlation matrix for the ASD-coADHD group. We can visually identify co-activation patterns highlighted in red along both the diagonals and the off-diagonals, forming a banded structure. By the ordering of ROIs in the 17-network parcellation, the first 500 ROIs correspond to the 17 networks of the left hemisphere, while the last 500 ROIs correspond to the 17 networks of the right hemisphere. Consequently, we can interpret the co-activation patterns along the diagonals as indicative of functional connectivity within each hemisphere, and the co-activation patterns along the off-diagonals as indicative of functional connectivity between the two hemispheres. Visual examination of the estimated correlation matrix shows that our proposed approach effectively captures the banded structure of the co-activation patterns observed in the sample correlation matrix. The plots for the ASD-nonADHD group are provided in Section 3 of the Supplementary Material.

\subsection{Interpreting functional connectivity between networks}

Our aim is to identify sets of positively correlated or co-activated ROIs that are specific to the ASD-coADHD group. To accomplish this, we compute a posterior similarity matrix where each entry represents the probability that pairs of ROIs belong to the same cluster. These probabilities are derived from empirically estimated posterior probabilities obtained through MCMC runs. See Section 3 of the Supplementary Material for the similarity matrices of the ASD-coADHD group and the ASD-nonADHD group. To achieve the highest possible accuracy in identifying correlated groups of ROIs, we focus on pairwise similarities where the similarity value is equal to or greater than 90\%. Subsequently, we exclude similarity pairs that appear in both the ASD-coADHD and ASD-nonADHD groups, as these pairs indicate connections shared by both groups. 

\begin{figure}
    \centering
    \includegraphics[width = 0.8\linewidth]{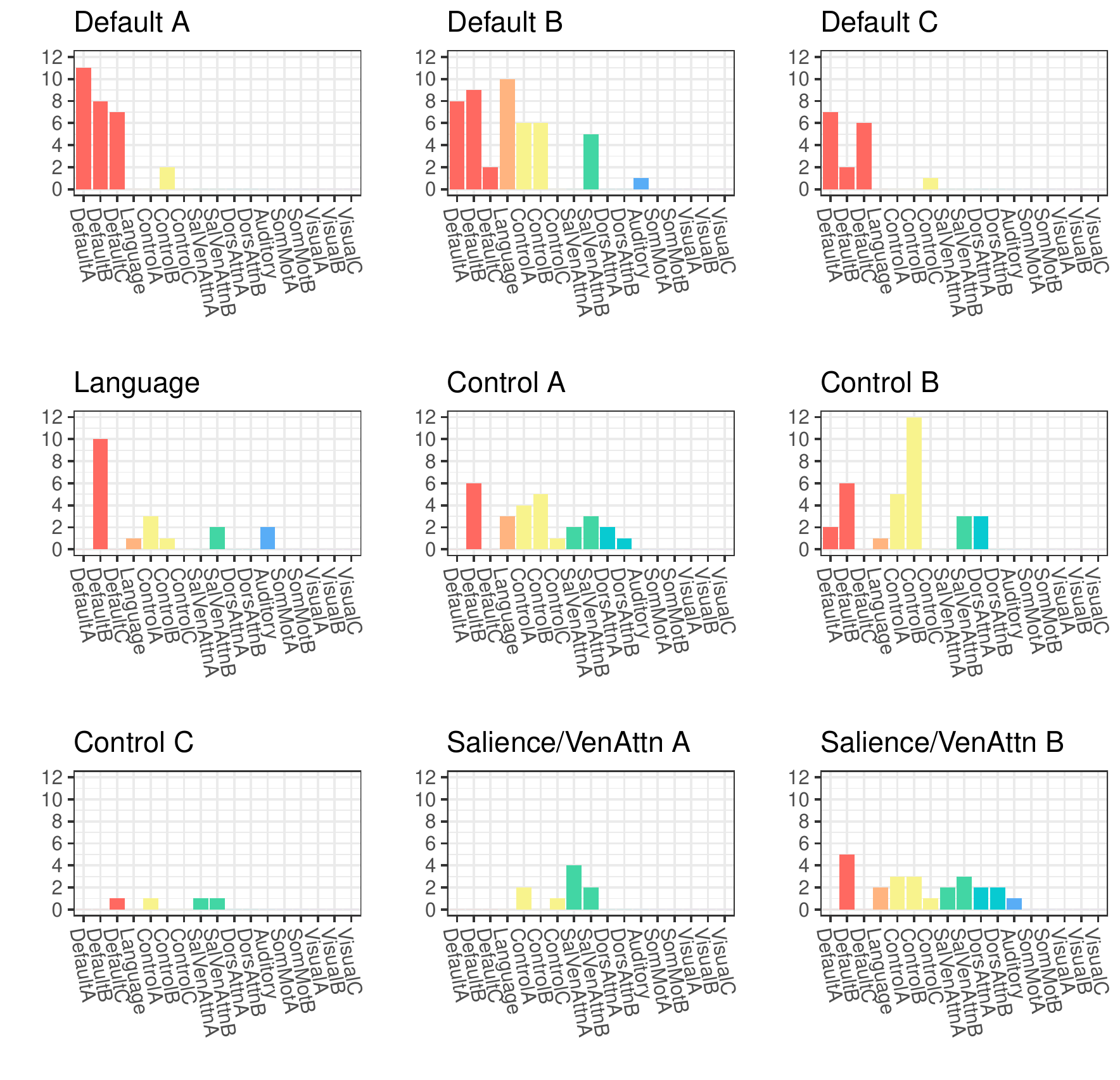}
    \caption{Summary of functional connectivity findings between the 17 networks. Plots of the number of ROIs connecting each network to the other 17 networks found in the ASD-coADHD group and not the ASD-nonADHD group.}
    \label{fig:networks}
\end{figure}

From the identified correlated groups of ROIs, we examine the functional connectivity between the 17 networks. The results of this analysis are summarized in Figure \ref{fig:networks}. In the figure, the number of ROIs connecting each network (9 out of 17) to the other 17 networks is presented. For the remaining 8 out of 17 networks, the results are detailed in Section 3 of the Supplementary Material.

Our findings are consistent with prior reports of multinetwork atypicalities in ADHD \citep{CastellanosAoki2016} and ASD \citep{Gaoetal2019, picci2016theoretical} that involve both subnetwork and internetwork connectivity. Unlike prior work primarily focusing on pairwise relationships between networks, our approach has the advantage of jointly identifying sets of interconnected networks and subnetworks. Results in the ASD-coADHD group pointed towards greater involvement of default subnetworks connectivity and their connections with the language network, as well as connectivity between the control and attention networks. These two patterns of findings have been independently reported in ASD \citep{Gaoetal2019, FarrantUddin2016} and ADHD \citep{Yerysetal2015, WangLiNiu2021}, suggesting that shared ASD- and ADHD-like patterns of atypical network connectivity exists in individuals with these frequent co-occurring conditions. Future studies with larger and well characterized samples of individuals with ASD, ADHD and their comorbidities are needed to confirm these findings.

\section{Conclusions} \label{sec:conc}

In this paper, we have introduced a Bayesian clustering approach that effectively groups covariance sub-matrices associated with high-dimensional Gaussian outcomes. Our method incorporates the stick-breaking construction of the Dirichlet process prior, enabling the identification of independent subgroups of outcomes. We proposed a new Head-Tail Split-Merge sampler to achieve convergence of the MCMC. Through extensive simulations, we have validated the performance of our approach in settings involving large dimensions of $M$ and $J$. Additionally, we have successfully applied our method to investigate co-activation patterns in ASD and ADHD, where we have discovered between-network functional connectivity patterns in the brain that align with findings in the existing literature.

In future research, it would be worthwhile to explore more comprehensive covariance structures to model functional connectivity in the brain for individuals with ASD. Possible directions include incorporating covariate effects into the covariance structure or considering ways to incorporate negative correlations in the covariance matrix. As observed in Figure \ref{fig:Cormat1000}, the sample correlation matrix exhibits not only co-activation patterns (positive correlation) but also reciprocal activation patterns (negative correlation). While the compound symmetry structure allows for negative correlations, its lower bound $-1/(M-1)$ tends to zero as $M$ increases, making it insufficient to capture reciprocal activation patterns. Further research in this area could explore separate modeling of both types of patterns.

\section*{Acknowledgements}

The authors are grateful to the participants of the ABIDE study, and the ABIDE study organizers and members who aggregated, preprocessed and shared the ABIDE data.

\bibliographystyle{plainnat}
\bibliography{references}

\end{document}


\title{\textbf{Supplementary Material} \\ \Large Bayesian estimation of clustered dependence structures in functional neuroconnectivity}
\author[1]{Hyoshin Kim}
\author[1]{Sujit K. Ghosh}
\author[2]{Adriana Di Martino}
\author[1]{Emily C. Hector \thanks{Hector was supported by a grant from the National Science Foundation (DMS2152887) and a Faculty Research and Professional Development Award from North Carolina State University.}}
\affil[1]{Department of Statistics, North Carolina State University}
\affil[2]{Autism Center, Child Mind Institute}
\date{}

\maketitle
\date{}

\section{Slice-within-Gibbs sampler for \texorpdfstring{$\boldsymbol{V}, \boldsymbol{\rho}, \boldsymbol{\sigma}^2, \alpha$, and $\boldsymbol{B}$}{Lg}} \label{sec:sampler}

This section provides a detailed explanation of the steps involved in running the Slice-within-Gibbs sampler to draw values from the conditional distributions of $\boldsymbol{V}, \boldsymbol{\rho}, \boldsymbol{\sigma}^2, \alpha$, and $\boldsymbol{B}$. The sampler iterates between the conditional distributions in the specified order. To commence, we introduce an expression for the likelihood which is used throughout the sampling algorithm.

\subsection{Expression for the Likelihood} \label{sec:Like}

We present an alternative representation of the likelihood: 
\begin{align}
    p \{\boldsymbol{Y}(\mathcal{S}) \,|\, \boldsymbol{B}, \boldsymbol{Z},\boldsymbol{\rho}, \boldsymbol{\sigma}^2 \} &\propto \text{det}\{ \boldsymbol{\Sigma}(\mathcal{S}, \mathcal{S}; \boldsymbol{\theta})\}^{-N/2} \label{eq:Like}\\
    &\,\,\,\,\,\,\,\, \cdot \exp \Bigl[ -\frac{1}{2} \sum_{i=1}^{N} \{\boldsymbol{y}_{i}(\mathcal{S}) - \boldsymbol{B}\boldsymbol{X}_{i}\}^{\top} \boldsymbol{\Sigma}(\mathcal{S}, \mathcal{S}; \boldsymbol{\theta})^{-1} \{\boldsymbol{y}_{i}(\mathcal{S}) - \boldsymbol{B}\boldsymbol{X}_{i}\} \Bigr]. \nonumber 
\end{align}
The determinant of $\boldsymbol{\Sigma}(\mathcal{S}, \mathcal{S}; \boldsymbol{\theta})$ in \eqref{eq:Like} can be expressed as 
\begin{align}
    &\label{eq:Det} \text{det}\{\boldsymbol{\Sigma}(\mathcal{S}, \mathcal{S}; \boldsymbol{\theta})\} \\ 
    &\propto \text{det}\{\boldsymbol{\Sigma}_{\text{perm}}(\mathcal{S}, \mathcal{S}; \boldsymbol{\theta})\} 
    = \prod_{j=1}^{J}  \text{det}\{\boldsymbol{\Sigma}_{\text{perm}}(\mathcal{S}_{j}, \mathcal{S}_{j}; \theta_{j})\} = \prod_{j=1}^{J} \sigma_{j}^{2d_{j}} \,\text{det}\{\boldsymbol{\Gamma}(\mathcal{S}_{j}, \mathcal{S}_{j}; \rho_{j})\}. \nonumber
\end{align}
Here, the terms involving the determinant of $\boldsymbol{P}_{\pi}$ can be treated as constants since the determinant of a permutation matrix is either $1$ or $-1$. Denoting $tr(\cdot)$ the trace of a matrix, the terms within the exponential in equation (\ref{eq:Like}) can be rewritten as 
\begin{align}
    &\sum_{i=1}^{N} \{\boldsymbol{y}_{i}(\mathcal{S}) - \boldsymbol{B}\boldsymbol{X}_{i}\}^{\top}  \bigl\{ \boldsymbol{P}_{\pi}^{\top} \boldsymbol{\Sigma}_{\text{perm}}(\mathcal{S}, \mathcal{S}; \boldsymbol{\theta}) \boldsymbol{P}_{\pi}\bigr\}^{-1} \{\boldsymbol{y}_{i}(\mathcal{S}) - \boldsymbol{B}\boldsymbol{X}_{i}\} \nonumber \\
    &= \sum_{i=1}^{N} \{\boldsymbol{y}_{i}(\mathcal{S}) - \boldsymbol{B}\boldsymbol{X}_{i} \}^{\top} \boldsymbol{P}_{\pi}^{\top} \boldsymbol{\Sigma}_{\text{perm}}^{-1}(\mathcal{S}, \mathcal{S}; \boldsymbol{\theta}) \boldsymbol{P}_{\pi}\{\boldsymbol{y}_{i}(\mathcal{S}) - \boldsymbol{B}\boldsymbol{X}_{i}\} \nonumber \\
    &= tr\bigl\{ \boldsymbol{A} \, \boldsymbol{\Sigma}_{\text{perm}}^{-1}(\mathcal{S}, \mathcal{S}; \boldsymbol{\theta}) \bigr\} \text{ where } \boldsymbol{A} = \boldsymbol{P}_{\pi} \Bigl[ \sum_{i=1}^{N} \{\boldsymbol{y}_{i}(\mathcal{S}) - \boldsymbol{B}\boldsymbol{X}_{i}\} \{\boldsymbol{y}_{i}(\mathcal{S}) - \boldsymbol{B}\boldsymbol{X}_{i} \}^{\top} \Bigr] \boldsymbol{P}_{\pi}^{\top} \label{eq:AGamma}\\ 
    &= \sum_{j=1}^{J} tr\bigl\{\boldsymbol{A}_{j} \boldsymbol{\Sigma}_{\text{perm}}^{-1}(\mathcal{S}_{j}, \mathcal{S}_{j}; \theta_{j})\bigr\} \text{ where } \boldsymbol{A}_{j} \in \mathbb{R}^{d_{j} \times d_{j}} \text{ is the } j^{th} \text{ diagonal block of }\boldsymbol{A} \nonumber\\
    &= \sum_{j=1}^{J} \sigma_{j}^{-2} \, tr\bigl\{\boldsymbol{A}_{j} \boldsymbol{\Gamma}^{-1}(\mathcal{S}_{j}, \mathcal{S}_{j}; \rho_{j})\bigr\} \nonumber.
\end{align}
The first equality follows from the fact that a permutation matrix is an orthogonal matrix. The computational challenges associated with calculating the expensive matrix inverse term $\boldsymbol{\Sigma}^{-1}$ can be avoided given the block-diagonal structure of $\boldsymbol{\Sigma}_{\text{perm}}^{-1}$. By combining terms in equations (\ref{eq:Det}) and (\ref{eq:AGamma}), the likelihood in equation (\ref{eq:Like}) can be expressed as: 
\begin{align*}
    p \{\boldsymbol{Y}(\mathcal{S}) \,|\, \boldsymbol{B},\boldsymbol{Z}, \boldsymbol{\rho}, \boldsymbol{\sigma}^2\} \propto \Bigl[\prod_{j=1}^{J} \sigma_{j}^{2d_{j}}  \text{det}\{\boldsymbol{\Gamma}(\mathcal{S}_{j}, \mathcal{S}_{j}; \rho_{j})\} \Bigr]^{-N/2} \exp\Bigl[-\frac{1}{2} \sum_{j=1}^{J} \sigma_{j}^{-2} \, tr \bigl\{\boldsymbol{A}_{j} \boldsymbol{\Gamma}^{-1}(\mathcal{S}_{j}, \mathcal{S}_{j}; \rho_{j}) \bigr\} \Bigr].
\end{align*}

\subsection{Conditional for \texorpdfstring{$\boldsymbol{V}$}{Lg}} \label{sec:SampleV}

By the conjugacy of the Beta distribution for $\boldsymbol{V}$ and the categorical distribution for $\boldsymbol{Z}$, the conditional posterior distribution for $\boldsymbol{V}$ is given by 
\begin{align*}
    p(\boldsymbol{V} \,|\, \cdot) &\propto p(\boldsymbol{Z} \,|\, \boldsymbol{V}) \,p(\boldsymbol{V} \,|\, \alpha) \\
    &\propto \prod_{j=1}^{K} \bigl\{w_{j}(\boldsymbol{V})^{m_{j}} \bigr\}
     \prod_{j=1}^{K-1} \bigl\{\alpha (1-V_{j})^{\alpha-1}\bigr\}
    \propto \prod_{j=1}^{K-1} \text{Beta}\Bigl(1+m_{j}, \alpha + \sum_{\ell=j+1}^{K}m_{\ell}\Bigr).
\end{align*}
Here, $m_{j}$ represents the count of instances where $Z_{m}=j$ for $m=1,\dots,M$, calculated as $m_{j} = \sum_{m=1}^{M}\mathbbm{1}(Z_{m} = j)$. To obtain posterior samples of $V_{j}$ for $j=1,\dots,K-1$, we independently draw from the following distribution (recalling $V_{K}=1$):
\begin{align*}
    p(V_{j} \,|\, \cdot) \propto \text{Beta}\Bigl(1+m_{j}, \alpha + \sum_{\ell=j+1}^{K}m_{\ell}\Bigr).
\end{align*}

\subsection{Conditional for \texorpdfstring{$\boldsymbol{\rho}$}{Lg}} \label{sec:Samplerho}

The conditional posterior distribution for $\boldsymbol{\rho}$ can be written as follows: 
\begin{align*}
    &p(\boldsymbol{\rho} \,|\, \cdot) \\
    &\propto p \bigl\{\boldsymbol{Y}(\mathcal{S}) \,|\, \boldsymbol{B}, \boldsymbol{Z}, \boldsymbol{\rho}, \boldsymbol{\sigma}^2\bigr\} \, p(\boldsymbol{\rho})\\
    &\propto \prod_{j=1}^{J} \Biggl[ \text{det} \bigl\{\boldsymbol{\Gamma}(\mathcal{S}_{j}, \mathcal{S}_{j}; \rho_{j})\bigr\}^{-\frac{N}{2}} \, \exp \Biggl[-\frac{1}{2 \sigma_{j}^{2}} \,  tr \bigl\{\boldsymbol{A}_{j} \boldsymbol{\Gamma}^{-1}(\mathcal{S}_{j}, \mathcal{S}_{j}; \rho_{j}) \bigr\} \Biggr] \, \eta_{j}^{a_2 - 1}(1-\eta_{j})^{b_2 - 1}  \Biggr] \\
    &\,\,\,\, \prod_{j=J+1}^{K} \left\{ \eta_{j}^{a_2 - 1}(1-\eta_{j})^{b_2 - 1}  \right\}.
\end{align*}
At each iteration, for $j=1,\dots,J$, we independently sample $\rho_{j}$ with
\begin{align*}
    p(\rho_{j} \,|\, \cdot) \propto \text{det} \bigl\{\boldsymbol{\Gamma}(\mathcal{S}_{j}, \mathcal{S}_{j}; \rho_{j})\bigr\}^{-\frac{N}{2}} \, \exp \Biggl[-\frac{1}{2 \sigma_{j}^{2}} \,  tr \bigl\{\boldsymbol{A}_{j} \boldsymbol{\Gamma}^{-1}(\mathcal{S}_{j}, \mathcal{S}_{j}; \rho_{j}) \bigr\} \Biggr] \, \eta_{j}^{a_2 - 1}(1-\eta_{j})^{b_2 - 1}.
\end{align*}
For $j=J+1,\dots,K-1$, we independently sample $\rho_{j}$ from
\begin{align*}
    p(\eta_j \,|\, \cdot) \propto \text{Beta}(a_{2}, b_{2}),
\end{align*}
where $\eta_{j}$ is the variable transformation of $\rho_{j}$. In the case where $d_{j} = 1$, we set $\rho_{j} = 0$ because the $j$-th sub-matrix has a dimension of one. This sampling procedure can be understood as follows: for $j=1,\dots,J$, where the cluster membership of the $j$-th block is updated, we sample $\rho_{j}$ by incorporating information from both the likelihood and the prior. However, for $j=J+1,\dots,K$, where the cluster membership of the $j$-th block remains unchanged, we update $\rho_{j}$ using only the prior information.

\begin{algorithm}[ht!]
\caption{Li and Walker's slice sampler \citep{LiWalker2020} for $\rho_{j}$}\label{alg:SSrho}
\begin{algorithmic}[1]

\REQUIRE Current values of $\rho_{0} := \rho_{j}$, step-size $s_{0} := s_{j}$; variance tuning parameter $\lambda$; support of $\rho_{j} \in (\rho_{\text{lb}}, \rho_{\text{ub}})$
\ENSURE Proposed values of $\rho_{j} := \rho_{1}$ and $s_{j} := s_{1}$
\STATE Sample the slice variable $\omega \sim U(0, p(\rho_{0} \,|\, \cdot))$, and sample 
\begin{align*}
    l_{1} \sim U(\rho_{0} - s_{0}/2, \rho_{0} + s_{0}/2) 
\end{align*} 
and $s_{1}$ from the density proportional to 
\begin{align*}
    \exp(-1/\lambda) \mathbbm{1}(s_{1} > 2 \lvert l_{1} - \rho_{0} \rvert)
\end{align*}
\STATE If $\rho_{\text{lb}} < l_{1}-s_{1}/2 < \rho_{\text{ub}}$ then set $a = l_{1}-s_{1}/2$ else $a = \rho_{\text{lb}}$ \\
If $\rho_{\text{lb}} < l_{1}+s_{1}/2 < \rho_{\text{ub}}$ then set $b = l_{1}+s_{1}/2$ else $b = \rho_{\text{ub}}$

\emph{}
\STATE Sample $\rho^{*} \sim U(a,b)$. If $p(\rho^{*} \,|\, \cdot) > \omega$, accept $\rho_{1} = \rho^{*}$. Else, 
\begin{align*}
    \text{if } \rho^{*} < \rho_{0}  \text{ then } a = \max(a, \rho^{*}) \text{ else } b = \min(b,\rho^{*})
\end{align*}
\STATE Repeat Step 3 until $p(\rho^{*} \,|\, \cdot) > \omega$ and set $\rho_{1} = \rho^{*}$
\end{algorithmic}
\end{algorithm}

We propose to independently sample $\rho_{j}$, $j=1,\dots,J$ in a Gibbs sampling framework using the slice sampler of \cite{LiWalker2020}. A single iteration of Li and Walker's slice sampler to generate a sample for $\rho_{j}$ is described in Algorithm \ref{alg:SSrho}. Slice sampling is a popular method for generating samples from complex target densities when direct sampling is difficult. Compared to the Metropolis-Hastings algorithm, slice samplers do not require manually specifying the proposal distribution and are known to exhibit good convergence properties. 

Li and Walker's slice sampler is an extension of Neal's slice sampler \citep{Neal2003}, with a modification made to the search component (Step 1 of Algorithm \ref{alg:SSrho}). The advantage of Li and Walker's slice sampler lies in its flexibility and simplicity. Neal's slice sampler uses a while-loop procedure with a fixed step size to search for the value of $\rho_{j}$ to sample. In contrast, Li and Walker's slice sampler adds flexibility by introducing a random step size $s_{j}$, eliminating the need for the while-loop component. The variance tuning parameter $\lambda$ for $s_{j}$ can be selected based on prior knowledge about the parameter's support, with Li and Walker recommending a value of either 10 or 100.

\subsection{Conditional for \texorpdfstring{$\boldsymbol{\sigma}^2$}{Lg}} \label{sec:Samplesigma2}

By the conjugacy of the Gaussian likelihood and the Inverse-Gamma prior distribution for $\boldsymbol{\sigma}^2$, the conditional posterior distribution for $\boldsymbol{\sigma}^2$ is given as
\begin{align*}
    &p(\boldsymbol{\sigma}^2 \,|\, \cdot) \\
    &\propto p\bigl\{\boldsymbol{Y}(\mathcal{S}) \,|\, \boldsymbol{B}, \boldsymbol{Z},\boldsymbol{\rho}, \boldsymbol{\sigma}^2 \bigr\} \, p(\boldsymbol{\sigma}^2)\\
    &\propto \prod_{j=1}^{J}  \text{Inv-Gamma}\Bigl\{ a_{1} + \frac{N d_{j}}{2}, \, b_{1} + \frac{1}{2} tr\bigl( \boldsymbol{A}_{j}\boldsymbol{\Gamma}^{-1}(\mathcal{S}_{j}, \mathcal{S}_{j}; \rho_{j})\bigr) \Big\} \prod_{j=J+1}^{K} \text{Inv-Gamma}(a_1,b_1).
\end{align*}
In a similar manner to sampling the correlation parameters, we independently sample $\sigma_{j}^2$ at each iteration for $j=1,\dots,J$ from
\begin{align*}
    p(\sigma_{j}^2 \,|\, \cdot) \propto \text{Inv-Gamma}\Bigl\{ a_{1} + \frac{N d_{j}}{2}, \, b_{1} + \frac{1}{2} tr\bigl( \boldsymbol{A}_{j}\boldsymbol{\Gamma}^{-1}(\mathcal{S}_{j}, \mathcal{S}_{j}; \rho_{j})\bigr) \Big\}.
\end{align*}
For $j=J+1,\dots,K-1$, we independently sample $\sigma_{j}^2$ from:
\begin{align*}
    p(\sigma_{j}^2\,|\, \cdot) \propto \text{Inv-Gamma}(a_1, b_1).
\end{align*}

\subsection{Conditional for \texorpdfstring{$\alpha$}{Lg}}

By the conjugacy of the Beta prior distribution for  $\boldsymbol{V}$ and the Gamma prior distribution for $\alpha$, we sample from the conditional posterior distribution for $\alpha$, given by
\begin{align*}
    p(\alpha\,|\, \cdot) &\propto p(\boldsymbol{V} \,|\, \alpha) \,p(\alpha) 
    \propto \text{Gamma}\Bigl\{a_{0}+K-1, b_{0} - \sum_{j=1}^{K-1}\log(1-V_{j}) \Bigr\}.
\end{align*}

\subsection{Conditional for \texorpdfstring{$\text{vec}(\boldsymbol{B})$}{Lg}} \label{sec:SampleB}

The form of the conditional posterior distribution for $\text{vec}(\boldsymbol{B})$ can be derived as follows. First, we re-express the mean model as:
\begin{align*}
    \boldsymbol{B}\boldsymbol{X}_{i}
    = \begin{Bmatrix}
    \text{diag}_{M}(X_{i1}) & \dots & \text{diag}_{M}(X_{ip})
    \end{Bmatrix} 
    \begin{pmatrix}
    \boldsymbol{\beta}_1\\
    \vdots\\
    \boldsymbol{\beta}_{p}
    \end{pmatrix}
    = \left(\boldsymbol{X}_{i}^{\top} \otimes \boldsymbol{I}_{M} \right)\text{vec}(\boldsymbol{B}^{})
    = \Tilde{\boldsymbol{X}}_{i}\text{vec}(\boldsymbol{B}).
\end{align*}
Here, $\text{diag}_{M}(X_{iq})$ represents a diagonal matrix of dimension $M$ with equal  diagonal entries $X_{iq}$, $\Tilde{\boldsymbol{X}}_{i} = (\boldsymbol{X}_{i}^{\top} \otimes \boldsymbol{I}_{M}) \in \mathbb{R}^{M \times Mp}$ and $\text{vec}(\boldsymbol{B}) \in \mathbb{R}^{Mp}$. To simplify notation, we denote $\boldsymbol{\Sigma} := \boldsymbol{\Sigma}(\mathcal{S},\mathcal{S}; \boldsymbol{\theta})$ and $\boldsymbol{\Sigma}_{\text{perm}} := \boldsymbol{\Sigma}_{\text{perm}}(\mathcal{S},\mathcal{S}; \boldsymbol{\theta})$. By the conjugacy of the Gaussian likelihood and the Gaussian prior distribution of $\text{vec}(\boldsymbol{B})$, the conditional posterior distribution for $\text{vec}(\boldsymbol{B})$ follows the multivariate Gaussian density
\begin{align*}
    &p\bigl\{\text{vec}(\boldsymbol{B}) \,|\, \cdot \bigr\}
\propto p \bigl\{\boldsymbol{Y}(\mathcal{S}) \,|\, \boldsymbol{B}, \boldsymbol{Z}, \boldsymbol{\rho}, \boldsymbol{\sigma}^2\bigr\}\,  p\bigl\{\text{vec}(\boldsymbol{B})\bigr\}
    \propto \mathcal{N}_{Mp} \left(\boldsymbol{\mu}, \boldsymbol{\Xi} \right), 
\end{align*}
where $\boldsymbol{\Xi}^{-1} := \boldsymbol{\Xi}^{-1}(\mathcal{S},\mathcal{S}; \boldsymbol{\theta})$ and $\boldsymbol{\mu} := \boldsymbol{\mu}(\mathcal{S}; \boldsymbol{\theta})$ are defined as
\begin{align*}
    \boldsymbol{\Xi}^{-1} = \sum_{i=1}^{N} \Tilde{\boldsymbol{X}}_{i}^{\top}\boldsymbol{\Sigma}^{-1} \Tilde{\boldsymbol{X}}_{i} + \tau^{-2} \boldsymbol{I}_{Mp}, \quad \boldsymbol{\mu} = \boldsymbol{\Xi} \sum_{i=1}^{N} \Tilde{\boldsymbol{X}}_{i}^{\top} \boldsymbol{\Sigma}^{-1} \boldsymbol{y}_{i}(\mathcal{S}).
\end{align*}
We further simplify the forms of $\boldsymbol{\Xi}^{-1}$, $\boldsymbol{\Xi}$ and $\boldsymbol{\mu}$ to avoid expensive matrix inverse calculations involving $\boldsymbol{\Sigma}^{-1}$. We can express $\boldsymbol{\Xi}^{-1}$ as a product of Kronecker products
\begin{align*}
    \boldsymbol{\Xi}^{-1} 
    &= (\boldsymbol{I}_{p} \otimes \boldsymbol{\Sigma}^{-1}) \sum_{i=1}^{N}   \Tilde{\boldsymbol{X}}_{i}^{\top} \Tilde{\boldsymbol{X}}_{i} + \tau^{-2}\boldsymbol{I}_{Mp}\\
    &= \bigl(\boldsymbol{I}_{p} \otimes \boldsymbol{\Sigma}^{-1} \bigr) \Biggl(\sum_{i=1}^{N}\boldsymbol{X}_{i}\boldsymbol{X}_{i}^{\top} \otimes \boldsymbol{I}_M \Biggr) + \tau^{-2}\boldsymbol{I}_{Mp}\\
    &= \left(\sum_{i=1}^{N}\boldsymbol{X}_{i}\boldsymbol{X}_{i}^{\top} \otimes (\boldsymbol{P}_{\boldsymbol{\pi}}^{\top}\boldsymbol{\Sigma}_{\text{perm}}^{-1}\boldsymbol{P}_{\boldsymbol{\pi}}) \right) + \tau^{-2}\boldsymbol{I}_{Mp}\\
    &=(\boldsymbol{I}_{p} \otimes \boldsymbol{P}_{\boldsymbol{\pi}}^{\top})\left( \sum_{i=1}^{N}\boldsymbol{X}_{i}\boldsymbol{X}_{i}^{\top} \otimes \boldsymbol{\Sigma}_{\text{perm}}^{-1} \right)(\boldsymbol{I}_{p} \otimes \boldsymbol{P}_{\boldsymbol{\pi}}) + \tau^{-2}(\boldsymbol{I}_{p} \otimes \boldsymbol{P}_{\boldsymbol{\pi}}^{\top})(\boldsymbol{I}_{p} \otimes \boldsymbol{P}_{\boldsymbol{\pi}})\\
    &= \bigl(\boldsymbol{I}_{p} \otimes \boldsymbol{P}_{\boldsymbol{\pi}}^{\top}\bigr)\Biggl( \sum_{i=1}^{N}\boldsymbol{X}_{i}\boldsymbol{X}_{i}^{\top} \otimes \boldsymbol{\Sigma}_{\text{perm}}^{-1} + \tau^{-2}\boldsymbol{I}_{Mp}\Biggr)\bigl(\boldsymbol{I}_{p} \otimes \boldsymbol{P}_{\pi} \bigr),
\end{align*}
where the first equality follows from noting that $\Tilde{\boldsymbol{X}}_i$ is a block matrix composed of row binding $\text{diag}_{M}(X_{iq})$, $q=1,\dots,p$. The eigendecomposition of the $p \times p$ real symmetric matrix $\sum_{i=1}^{N}\boldsymbol{X}_{i}\boldsymbol{X}_{i}^{\top}$ and the $M \times M$ real symmetric matrix $\boldsymbol{\Sigma}_{\text{perm}}^{-1}$ yields $\sum_{i=1}^{N}\boldsymbol{X}_{i}\boldsymbol{X}_{i}^{\top} = \boldsymbol{Q}_1 \boldsymbol{\Lambda}_1 \boldsymbol{Q}_1^{\top}$ and $\boldsymbol{\Sigma}_{\text{perm}}^{-1} = \boldsymbol{Q}_2 \boldsymbol{\Lambda}_2 \boldsymbol{Q}_2^{\top}$, respectively. Here, $\boldsymbol{\Lambda}_1$ and $\boldsymbol{\Lambda}_2$ are diagonal matrices, and $\boldsymbol{Q}_1$ and $\boldsymbol{Q}_2$ are orthogonal matrices. Substituting these expressions into $\boldsymbol{\Xi}$ yields
\begin{align}
    \boldsymbol{\Xi}
    &= (\boldsymbol{I}_{p} \otimes \boldsymbol{P}_{\boldsymbol{\pi}}^{\top})\left( \sum_{i=1}^{N}\boldsymbol{X}_{i}\boldsymbol{X}_{i}^{\top} \otimes \boldsymbol{\Sigma}_{\text{perm}}^{-1} + \tau^{-2}I_{Mp}\right)^{-1}(\boldsymbol{I}_{p} \otimes \boldsymbol{P}_{\boldsymbol{\pi}}) \nonumber\\
    &= (\boldsymbol{I}_{p} \otimes \boldsymbol{P}_{\boldsymbol{\pi}}^{\top}) (\boldsymbol{Q}_1 \otimes \boldsymbol{Q}_2)(\boldsymbol{\Lambda}_1 \otimes \boldsymbol{\Lambda}_2 + \tau^{-2}I_{Mp})^{-1}(\boldsymbol{Q}_1 \otimes \boldsymbol{Q}_2)^{\top} (\boldsymbol{I}_{p} \otimes \boldsymbol{P}_{\boldsymbol{\pi}}) \label{eq:U} \\
    &= (\boldsymbol{Q}_1 \otimes \boldsymbol{P}_{\boldsymbol{\pi}}^{\top} \boldsymbol{Q}_2) (\boldsymbol{\Lambda}_1 \otimes \boldsymbol{\Lambda}_2 + \tau^{-2}I_{Mp})^{-1}(\boldsymbol{Q}_1 \otimes \boldsymbol{P}_{\boldsymbol{\pi}}^{\top} \boldsymbol{Q}_2)^{\top} \nonumber \\
    &= \boldsymbol{U} \boldsymbol{U}^{\top} \nonumber
\end{align}
where $\boldsymbol{U} := (\boldsymbol{Q}_1 \otimes \boldsymbol{P}_{\pi}^{\top} \boldsymbol{Q}_2) (\boldsymbol{\Lambda}_1 \otimes \boldsymbol{\Lambda}_2 + \tau^{-2}\boldsymbol{I}_{Mp})^{-1/2}$. 
This reduces the computationally intensive inverse of a $Mp \times Mp$ matrix to a computation of the inverse square root of the diagonal matrix $\boldsymbol{U}$. Since $\boldsymbol{\Sigma}_{\text{perm}}^{-1}$ is a block diagonal matrix, we can reduce computational time by separately performing the eigendecomposition of each main diagonal block to obtain $\boldsymbol{Q}_2$ and $\boldsymbol{\Lambda}_{2}$. The eigendecomposition for $\sum_{i=1}^{N}\boldsymbol{X}_{i}\boldsymbol{X}_{i}^{\top}$ is performed once at the beginning of the sampling algorithm. Similarly, we express $\boldsymbol{\mu}$ as follows:
\begin{eqnarray}
\boldsymbol{\mu} 
    &=& \boldsymbol{\Xi} \, \sum_{i=1}^{N} (\boldsymbol{I}_{p} \otimes  \boldsymbol{\Sigma}^{-1}) (\boldsymbol{X}_i \otimes  \boldsymbol{I}_M ) \boldsymbol{y}_i(\mathcal{S}) \nonumber \\
    &=& \boldsymbol{\Xi} \, \sum_{i=1}^{N} (\boldsymbol{X}_i \otimes \boldsymbol{\Sigma}^{-1} )\boldsymbol{y}_i(\mathcal{S}) \label{eq:mu} \\
    &=&\boldsymbol{\Xi} \, \text{vec}\Bigl \{ \boldsymbol{P}_{\pi}^{\top}\boldsymbol{\Sigma}_{\text{perm}}^{-1}\boldsymbol{P}_{\pi} \sum_{i=1}^{N}\boldsymbol{y}_i(\mathcal{S}) \boldsymbol{X}_i^{\top} \Bigr \}. \nonumber
\end{eqnarray}

Here, the last equality follows from the vectorization of the Kronecker product. By combining the terms in equations (\ref{eq:U}) and (\ref{eq:mu}), we sample $\text{vec}(\boldsymbol{B})$ from the following distribution:
\begin{align*}
    \text{vec}(\boldsymbol{B}) = \boldsymbol{\mu} + \boldsymbol{U}\boldsymbol{\nu}, \quad \boldsymbol{\nu} \sim \mathcal{N}_{Mp} \left(\boldsymbol{0}, \boldsymbol{I}_{Mp} \right).
\end{align*}


\section{Additional simulation outputs}
\subsection{First simulation}

\begin{figure}[ht!]
\centering
\includegraphics[width=1\textwidth]{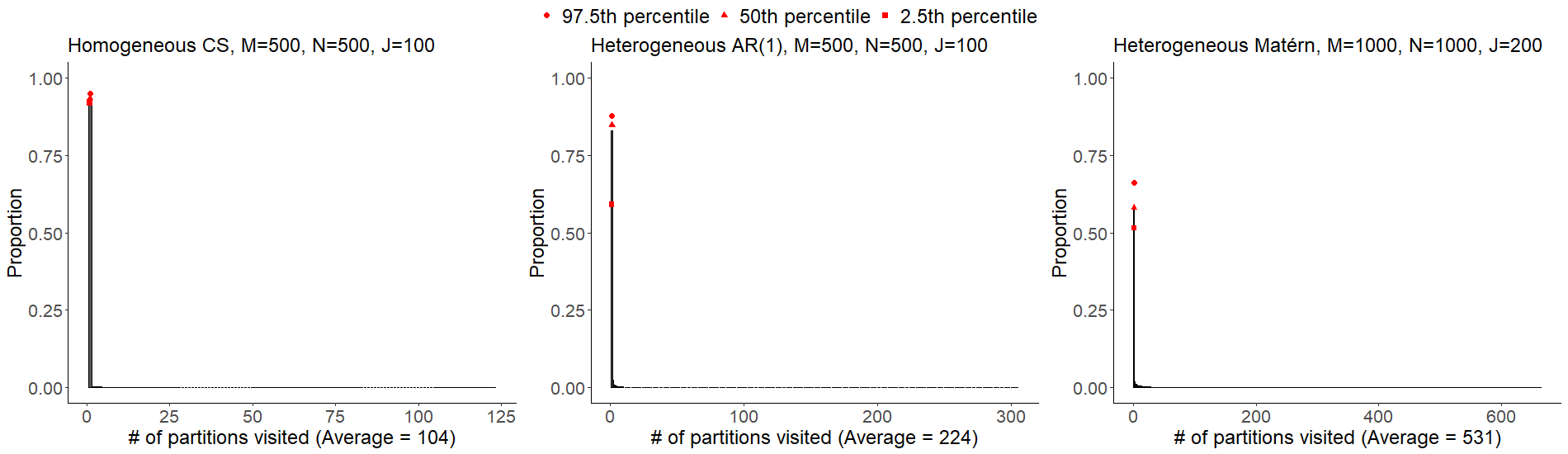}
\caption{First simulation -- Plot of posterior distribution of $\boldsymbol{Z}$ averaged across 50 simulations during the burn-in period, assuming a uniform true partition. The red points are the 95\% percentile intervals of the probability of MAPs across the 50 simulations.}
\label{fig:Zplot_uniform_burnin}
\end{figure}

This section presents the results from the first simulation during the burn-in period. Figure \ref{fig:Zplot_uniform_burnin} displays the average posterior distribution of $\boldsymbol{Z}$ across 50 simulations for the three settings. As the complexity of the covariance structure increases, the average number of partitions visited also increases, while the posterior probability of the MAP decreases. This suggests that a greater number of iterations is needed to identify the true partition as the complexity increases. In addition, the 95\% percentile intervals of the posterior probability of the MAP derived from the 50 simulations become wider with increasing complexity, indicating that some simulations fail to accurately identify the correct partition during the burn-in period. 

Table \ref{table:CR_uniform_burnin} presents the average coverage rates for a 95\% credible interval across 50 simulations for $\boldsymbol{\rho}, \boldsymbol{\sigma}^2$, and $\boldsymbol{B}$ during the burn-in period. The coverage rates are computed from the subset of iterations where $\boldsymbol{Z}$ is equal to its MAP estimate. The results demonstrate that the desired 95\% level is achieved for all three settings, with slightly conservative inference and greater variability compared to the sampling period.

\begin{table}[ht!]
\caption{First simulation -- Summary of coverage rates for a 95\% credible interval, averaged across 50 simulations for $\boldsymbol{\rho}, \boldsymbol{\sigma}^2, \boldsymbol{B}$ during the burn-in period, assuming a uniform true partition. Standard deviations are in parentheses. The coverage rates are computed from the subset of iterations where $\boldsymbol{Z}$ is equal to its MAP estimate.}
\centering
\begin{tabular}{lcccccc}
& & & & \multicolumn{3}{c}{coverage rate (sd)}\\
\cline{5-7}
                      & $M$ & $N$ & $J$ & $\boldsymbol{\rho}$ & $\boldsymbol{\sigma}^2$ & $\boldsymbol{B}$ \\ \hline
homogeneous CS        & 500 & 500 & 100 & 1.00 (0.00) & 1.00 (0.00) & 0.95 (0.01) \\
heterogeneous AR(1)     & 500 & 500 & 100 & 0.96 (0.07) & 0.97 (0.05) & 0.95 (0.01) \\
heterogeneous Mat\'ern  & 1000 & 1000 & 200 & 0.96 (0.14) & 0.97 (0.08) & 0.95 (0.01) \\ \hline
\end{tabular}
\label{table:CR_uniform_burnin}
\end{table}

\subsection{Second simulation} \label{sec:add_sim}

Here we present results from the second simulation during the burn-in period. 

\begin{figure}[ht!]
\centering
\includegraphics[width=0.6\textwidth]{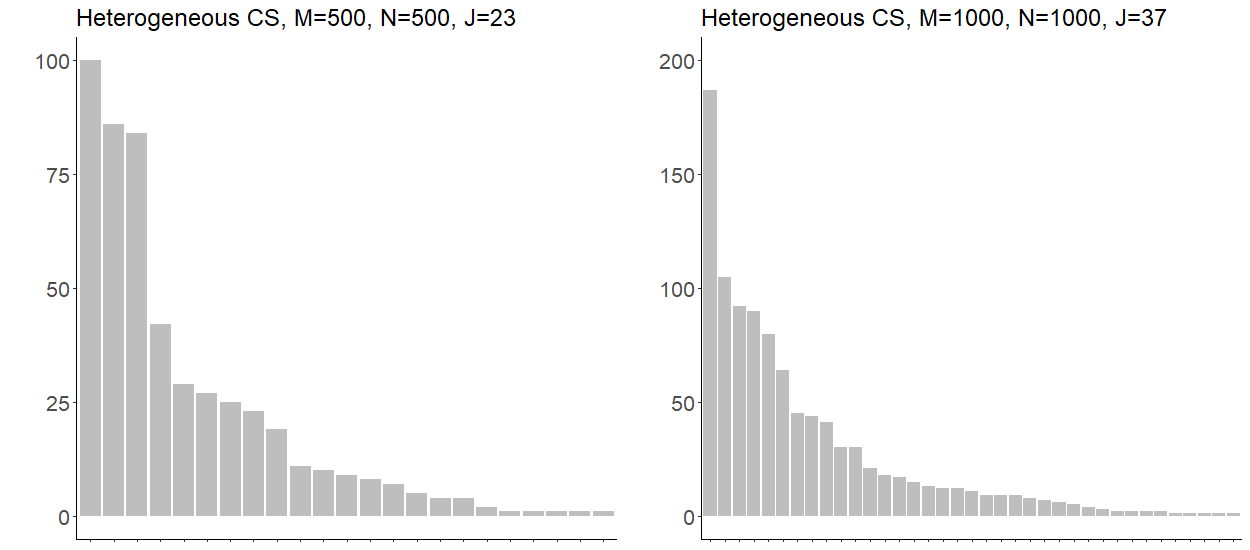}
\caption{Second simulation -- Plot of the true partition generated from a $\text{DP}(6,P_{0})$. The partition labels are sorted by block size from largest to smallest.}
\label{fig:Hist_DP}
\end{figure}

\begin{figure}[ht!]
\centering
\includegraphics[width=0.7\textwidth]{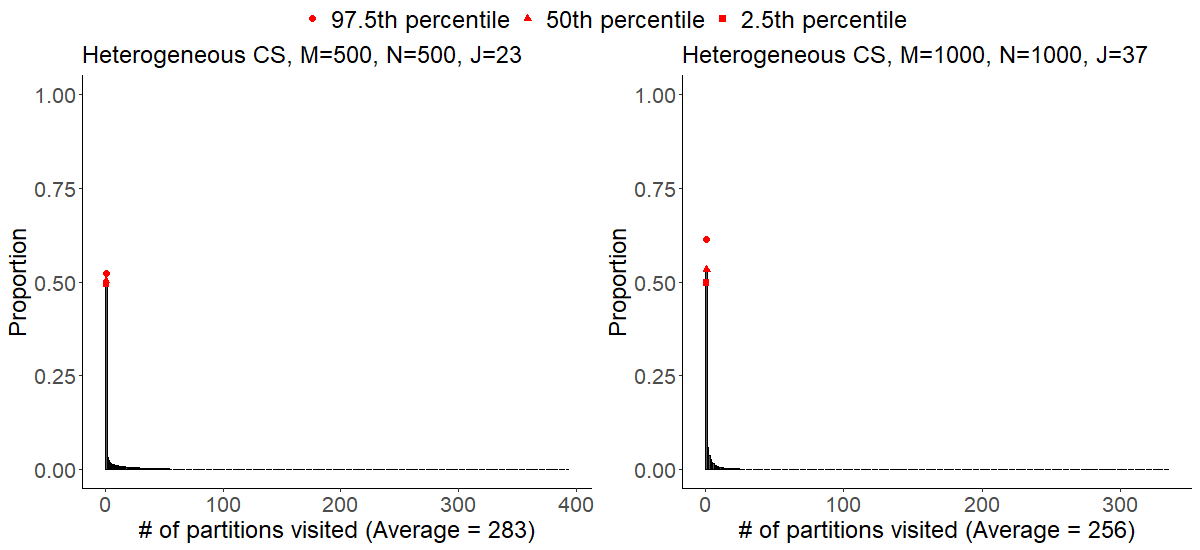}
\caption{Second simulation -- Plot of posterior distribution of $\boldsymbol{Z}$ averaged across 50 simulations during the burn-in period, assuming a $\text{DP}(6,P_{0})$ true partition. The red points are the 95\% percentile intervals of the probability of MAPs across the 50 simulations.}
\label{fig:Zplot_DP_burnin}
\end{figure}

Figure \ref{fig:Zplot_DP_burnin} shows that the average probability assigned to the MAP estimate is approximately 0.5 for both settings during the burn-in period, due to the proposed HTSM algorithm's tail moves during the second phase of the MCMC. As the true partition has a moderate amount of size one clusters, the sampler mostly identifies the true partition during the second phase. Note that in the first simulation, the true partition was mostly identified during the first phase, leading to higher posterior probability of the MAPs than the second simulation. 

Table \ref{table:CR_DP_burnin} shows that the desired 95\% level is achieved for the average coverage rates of credible intervals across 50 simulations for $\boldsymbol{\rho}, \boldsymbol{\sigma}^2$, and $\boldsymbol{B}$ during the burn-in period, with greater variability compared to the sampling phase. 

\begin{table}[ht!]
\caption{Second simulation -- Summary of coverage rates for a 95\% credible interval, averaged across 50 simulations for $\boldsymbol{\rho}, \boldsymbol{\sigma}^2, \boldsymbol{B}$ during the burn-in period, assuming a $\text{DP}(6,P_{0})$ true partition. Standard deviations are in parentheses. The coverage rates are computed from the subset of iterations where $\boldsymbol{Z}$ is equal to its MAP estimate.
}
\centering
\begin{tabular}{lcccccc}
& & & & \multicolumn{3}{c}{coverage rate (sd)}\\
\cline{5-7}
                      & $M$ & $N$ & $J$ & $\boldsymbol{\rho}$ & $\boldsymbol{\sigma}^2$ & $\boldsymbol{B}$ \\ \hline
heterogeneous CS      & 500 & 500 & 23 & 0.99 (0.02) & 0.97 (0.03) & 0.95 (0.02) \\
heterogeneous CS      & 1000 & 1000 & 37 & 0.99 (0.02) & 0.97 (0.03) & 0.95 (0.01) \\ \hline
\end{tabular}
\label{table:CR_DP_burnin}
\end{table}


\newpage

\section{Additional data analysis outputs} \label{sec:add_dat}

\begin{figure}[ht!]
\centering
\includegraphics[width=1\textwidth]{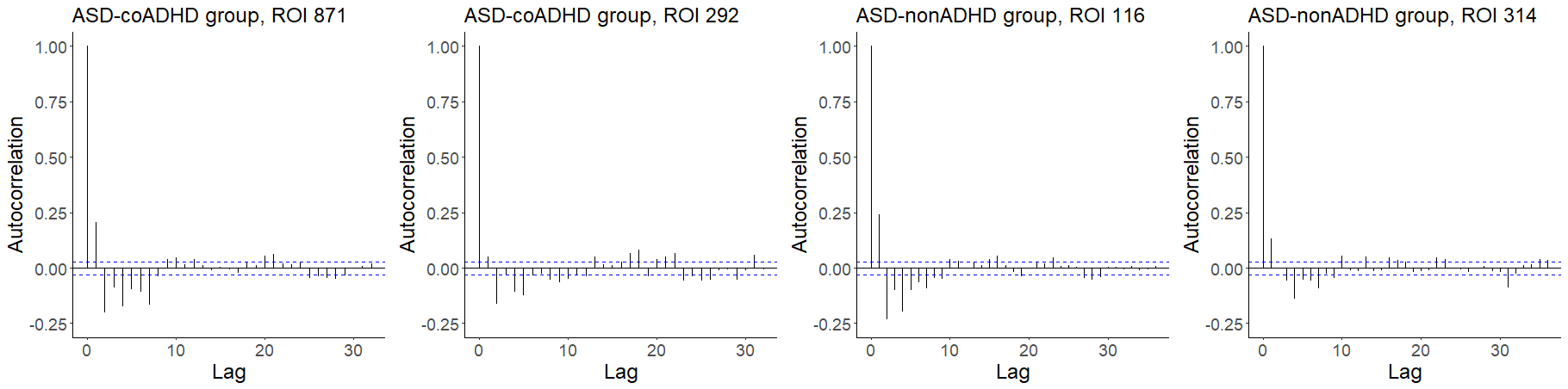}
\caption{Autocorrelation functions of the rfMRI time series with a lag of 2 for the ASD-coADHD group and the ASD-nonADHD group, on randomly selected ROIs.}
\label{fig:acf}
\end{figure}

This section presents additional data analysis outputs. Figure \ref{fig:acf} shows that the autocorrelation of the rfMRI time series outcomes diminishes to zero, indicating that the outcomes can be treated as independent.

\begin{figure}[ht!]
\centering
\includegraphics[width=0.9\textwidth]{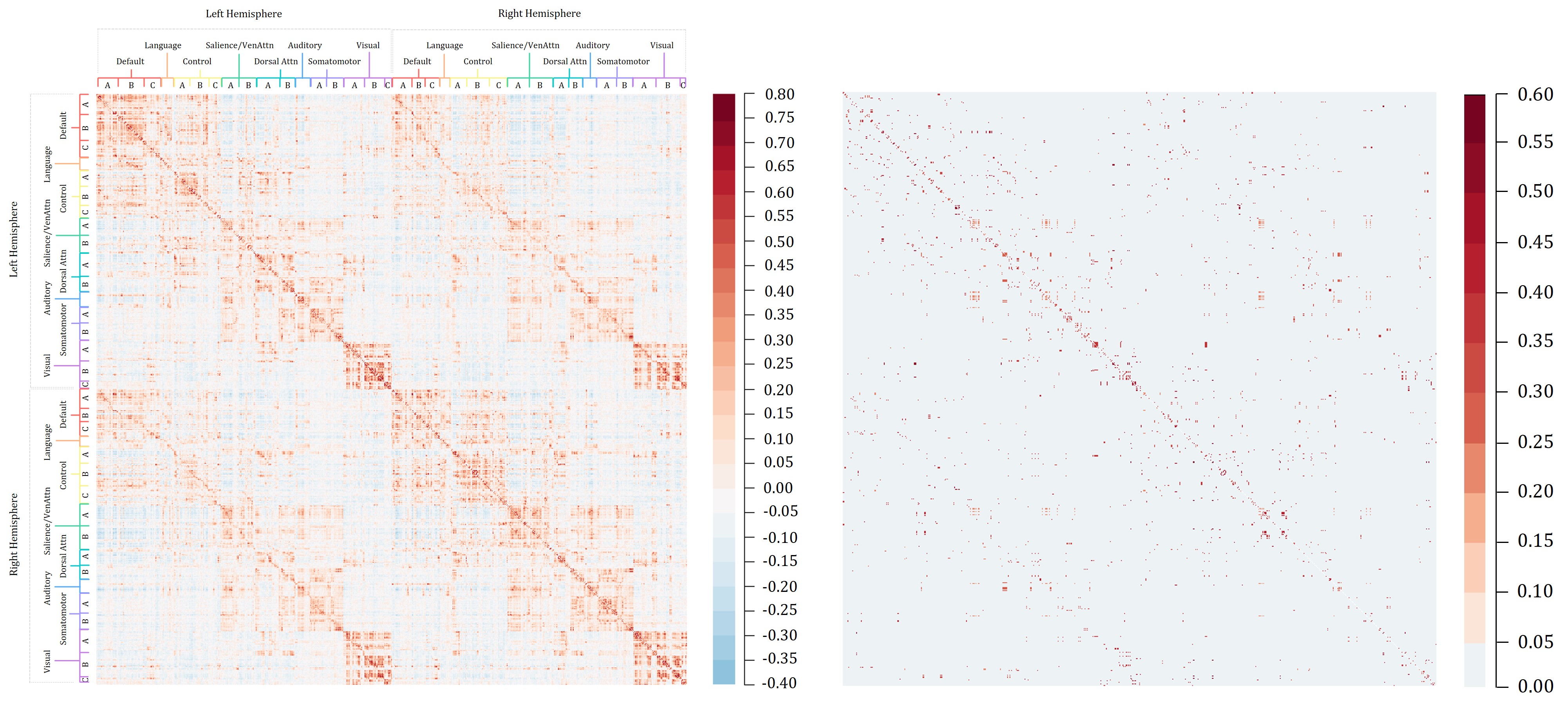}
\caption{Connectivity matrix (left) and the estimated correlation matrix (right) for the ASD-nonADHD group. The initial 500 ROIs correspond to the 17 networks of the left hemisphere, while the latter 500 ROIs correspond to the 17 networks of the right hemisphere.}
\label{fig:Cormat1000_nocomorbid}
\end{figure}

\begin{figure}[ht!]
    \centering
    \includegraphics[width = 0.7\linewidth]{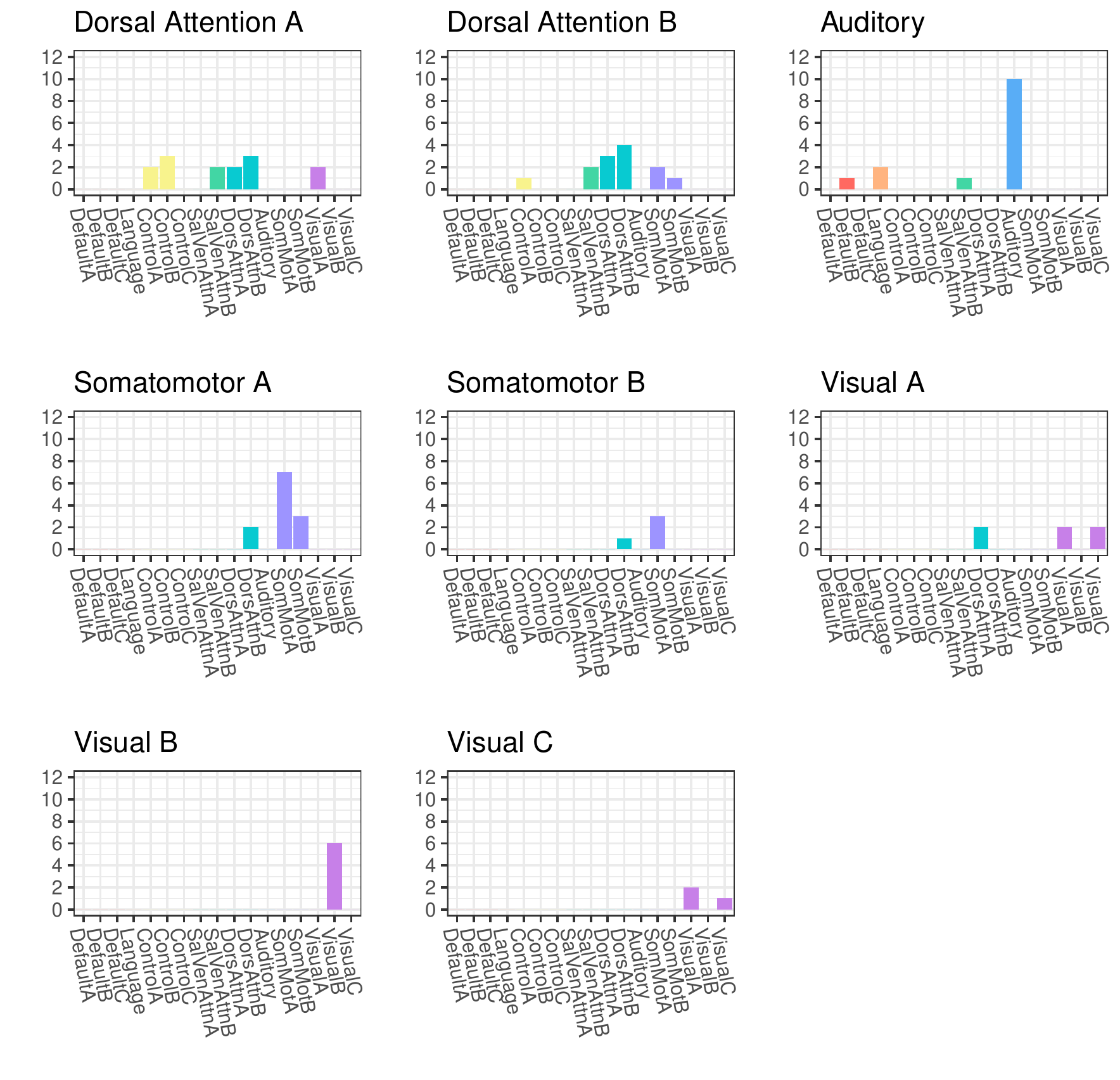}
    \caption{Summary of functional connectivity findings between the 17 networks. Plots of the number of ROIs connecting each network to the other 17 networks found in the ASD-coADHD group and not the ASD-nonADHD group.}
    \label{fig:networks}
\end{figure}

\begin{figure}[ht!]
\centering
\includegraphics[width=0.9\textwidth]{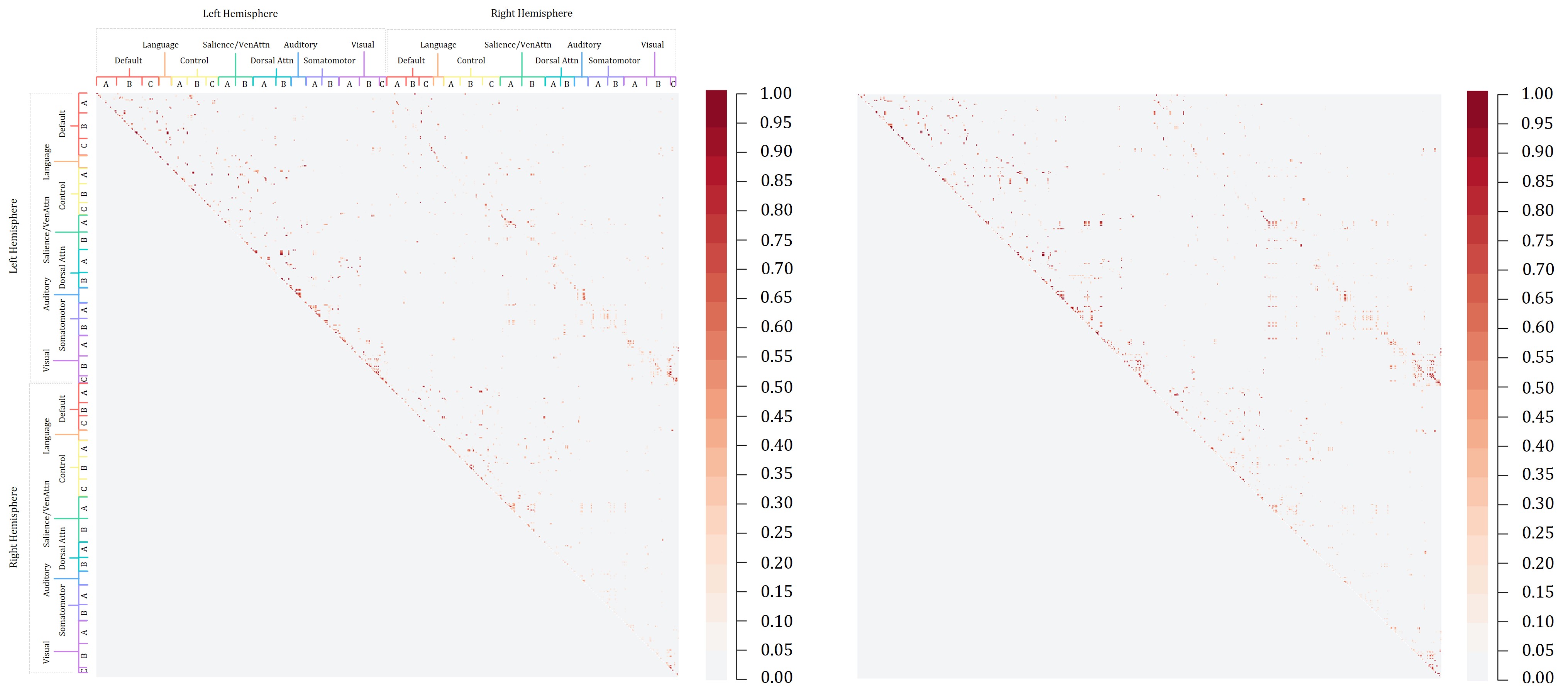}
\caption{Posterior similarity matrix indicating the probability that pairs of ROIs belong to the same cluster for the ASD-coADHD group (left) and for the ASD-nonADHD group (right). }
\label{fig:Simmat1000}
\end{figure}

\newpage
\bibliographystyle{plainnat}
\bibliography{references}       